\documentclass[letterpaper]{article} 
\usepackage{aaai2026}  
\usepackage{times}  
\usepackage{helvet}  
\usepackage{courier}  
\usepackage[hyphens]{url}  
\usepackage{graphicx} 
\usepackage{natbib}  
\usepackage{caption} 
\usepackage{algorithm}
\usepackage{algorithmic}

\usepackage{booktabs} 
\usepackage{multirow}
\usepackage{graphicx}
\usepackage{adjustbox}
\usepackage{makecell}  
\usepackage{amssymb}

\usepackage{newfloat}
\usepackage{listings}
\DeclareCaptionStyle{ruled}{labelfont=normalfont,labelsep=colon,strut=off} 
\floatstyle{ruled}
\newfloat{listing}{tb}{lst}{}
\floatname{listing}{Listing}
\nocopyright

\title{OSUM-EChat: Enhancing End-to-End Empathetic Spoken Chatbot via Understanding-Driven Spoken Dialogue}
\author{
    Xuelong Geng\textsuperscript{\rm 1}\thanks{These authors contributed equally as first authors.},
    Qijie Shao\textsuperscript{\rm 1}\footnotemark[1],
    Hongfei Xue\textsuperscript{\rm 1}\footnotemark[1],
    Shuiyuan Wang\textsuperscript{\rm 1}\thanks{These authors contributed equally as second authors.},
    Hanke Xie\textsuperscript{\rm 1}\footnotemark[2],
    Zhao Guo\textsuperscript{\rm 1},
    Guojian Li\textsuperscript{\rm 1},
    Wenjie Tian\textsuperscript{\rm 1},
    Chengyou Wang\textsuperscript{\rm 1},
    Zhixian Zhao\textsuperscript{\rm 1},
    Kangxiang Xia\textsuperscript{\rm 1},
    Ziyu Zhang\textsuperscript{\rm 1},
    Zhennan Lin\textsuperscript{\rm 1},
    Tianlun Zuo\textsuperscript{\rm 1},
    Mingchen Shao\textsuperscript{\rm 1},
    Yuang Cao\textsuperscript{\rm 1},
    Guobin Ma\textsuperscript{\rm 1},
    Longhao Li\textsuperscript{\rm 1},
    Yuhang Dai\textsuperscript{\rm 1},
    Dehui Gao\textsuperscript{\rm 1},
    Dake Guo\textsuperscript{\rm 1},
    Lei Xie\textsuperscript{\rm 1}\thanks{Corresponding author.}
}
\affiliations{
    \textsuperscript{\rm 1}Audio, Speech and Language Processing Group (ASLP@NPU),\\
Northwestern Polytechnical University, Xi’an\\
    xlgeng@mail.nwpu.edu.cn, lxie@nwpu.edu.cn\\
    \vspace{0.5em} 
    \centering
    \textbf{Project page:} \url{https://github.com/ASLP-lab/OSUM}
}

\usepackage{bibentry}

\begin{document}

\maketitle

\begin{abstract}
Empathy is crucial in enabling natural interactions within spoken dialogue systems, allowing machines to recognize and respond appropriately to paralinguistic cues such as age, gender, and emotion. Recent advancements in end-to-end speech language models, which unify speech understanding and generation, provide promising solutions. However, several challenges persist, including an over-reliance on large-scale dialogue datasets, insufficient extraction of paralinguistic cues vital for conveying empathy, and the lack of empathy-specific datasets and evaluation frameworks. To address these issues, we introduce OSUM-EChat, an open-source, end-to-end spoken dialogue system designed to enhance empathetic interactions, particularly in resource-limited settings. OSUM-EChat introduces two key innovations: (1) a three-stage \textit{understanding-driven spoken dialogue} training strategy that extends the capabilities of a large speech understanding model to spoken dialogue tasks, and (2) a \textit{linguistic–paralinguistic dual thinking} mechanism that integrates paralinguistic understanding through a chain of thought with dialogue generation, enabling the system to produce more empathetic responses. This approach reduces reliance on large-scale dialogue datasets while maintaining high-quality empathetic interactions. Additionally, we introduce the \textit{EChat-200K} dataset, a rich corpus of empathetic speech-to-speech dialogues, and the \textit{EChat-eval} benchmark, a comprehensive framework for evaluating the empathetic capabilities of dialogue systems. Experimental results demonstrate that OSUM-EChat outperforms end-to-end spoken dialogue models regarding empathetic responsiveness, validating its effectiveness. 

\end{abstract}


\section{Introduction}
\begin{figure}[t]
    \centering
    \includegraphics[width=1\linewidth]{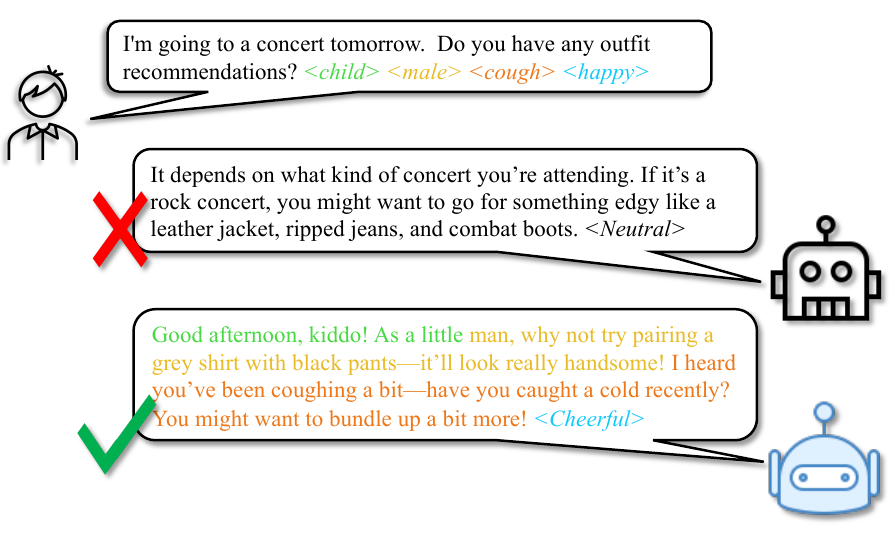}
    \caption{Empathetic spoken dialogue scenario. The model responds to the paralinguistic cues in angle brackets.}
    \label{fig: OSUM-EChat-empathetic}
\end{figure}

Empathy is crucial for achieving natural interactions in spoken dialogue systems, as it enables the perception and response to paralinguistic cues in speech—such as age, gender, and emotion.
In human communication, empathy plays a crucial role in fostering meaningful connections by enabling individuals to recognize emotional nuances and respond appropriately. For machines, achieving this level of sensitivity is particularly challenging but essential for generating authentic and emotionally intelligent responses. As shown in Figure~\ref{fig: OSUM-EChat-empathetic}, a user's speech query may carry rich paralinguistic cues, such as age (\textit{child}), gender (\textit{male}), caption (\textit{cough}), or emotion (\textit{happy}). An empathetic response must not only address the linguistic core of the query, e.g., outfit recommendations, but also align with these cues by adopting an appropriate tone and using language suited to the context, e.g., a cheerful tone and child-friendly vocabulary.
However, typical cascaded dialogue systems, which sequentially process speech input through Automatic Speech Recognition (ASR), Large Language Models (LLMs), and Text-to-Speech (TTS), inherently strip away critical paralinguistic information. To address this limitation, end-to-end speech language models have emerged, unifying speech encoding and generation into a single framework to better preserve this information~\cite{gpt4o}.

End-to-end spoken dialogue systems can generally be categorized into two types~\cite{s2s:align_minmo}: modular aligned multimodal models and native multimodal models.
The former, exemplified by Freeze-Omni~\cite{s2s:align_freeze_omni} and Qwen2.5-Omni~\cite{s2s:align_qwen2.5_omni}, integrate speech encoders and decoders with LLMs to handle speech understanding and generation tasks independently. While this approach leverages the general-purpose capabilities of LLMs, it struggles to capture and generate fine-grained paralinguistic cues because it models the two modalities independently. 
In contrast, native multimodal models, such as Mini-Omni~\cite{s2s:full_mini_omni} and GLM-4-Voice~\cite{s2s:full_glm4voice}, integrate the speech modality directly into the LLM architecture. By allowing the LLM to model speech signals natively, these models are better equipped to retain crucial paralinguistic cues~\cite{s2s:full_deeptalk}, vital for generating empathetic dialogue.

Despite recent advancements in end-to-end spoken dialogue systems, native multimodal models still face substantial challenges in empathetic dialogue, particularly in three key aspects.
First, closed-source proprietary models over-rely on huge, diverse spoken dialogue data to implicitly capture paralinguistic cues~\cite{gpt4o}. This creates a critical barrier in resource-limited settings, where such data is scarce.
Second, current open-source models often rely on linguistically annotated data, such as ASR transcripts, for speech modality alignment~\cite{s2s:full_baichuan_audio, s2s:full_step_audio, s2s:full_glm4voice}, which leads to suboptimal paralinguistic modeling and thereby weakens the consistency of empathetic responses.
Finally, the scarcity of high-quality empathetic dialogue data and empathy-focused evaluation benchmarks exacerbates the limitations above, further hindering the effective modeling and quantitative assessment of paralinguistic cues in spoken dialogue systems.

To address these challenges, this paper introduces \textbf{OSUM-EChat}, an end-to-end empathetic spoken dialogue system that enhances paralinguistic modeling despite data constraints. Specifically, we introduce an \textit{understanding-driven spoken dialogue} training strategy that encompasses developing a multitask speech understanding model grounded in the OSUM framework~\cite{osum}, further extending its speech generation capabilities to support speech-to-speech dialogue. Within this strategy, a \textit{linguistic-paralinguistic dual think} mechanism is purposefully designed to transfer paralinguistic understanding capabilities to dialogue systems better, thereby forming a comprehensive end-to-end system that enhances paralinguistic modeling while minimizing reliance on large-scale data and computational resources.
To address the lack of empathetic conversation data and evaluation benchmarks, we introduce the \textit{EChat-200K} dataset and the \textit{EChat-eval} benchmark. The \textit{EChat-200K} dataset is a speech-to-speech dialogue corpus rich in paralinguistic information. The \textit{EChat-eval} benchmark is a comprehensive evaluation designed to assess the empathetic capabilities of dialogues across multiple paralinguistic dimensions.
Experimental results demonstrate that OSUM-EChat outperforms previous end-to-end spoken models regarding empathetic responses, validating its effectiveness in resource-limited empathetic dialogue tasks.

The key contributions are summarized as follows:
\begin{itemize}
\item We introduce OSUM-EChat, an end-to-end empathetic spoken dialogue model that demonstrates strong performance to perceive users' paralinguistic cues and generate contextually aligned empathetic responses, as validated on benchmark datasets.

\item We present an \textit{understanding-driven spoken dialogue} training strategy and a \textit{linguistic-paralinguistic dual think} reasoning mechanism to facilitate knowledge transfer from speech understanding models to spoken dialogue tasks, enabling empathetic dialogue modeling in resource-limited settings.

\item We introduce the \textit{EChat-200K} dataset and \textit{EChat-eval} benchmark. The \textit{EChat-200K} is a speech-to-speech empathetic dialogue corpus rich in comprehensive cues, while the \textit{EChat-eval} serves as a benchmark for assessing empathetic responses across multiple paralinguistic dimensions, addressing the gap in current evaluations that lack comprehensive multi-dimensional frameworks.

\item We provide a complete open-source ecosystem for empathetic dialogue research, including the model, whole codebases (training, inference, deployment), datasets, and data construction workflows to foster community-driven advancements. Samples can be found.
\end{itemize}

\begin{figure}[t]
    \centering
    \includegraphics[width=\linewidth]{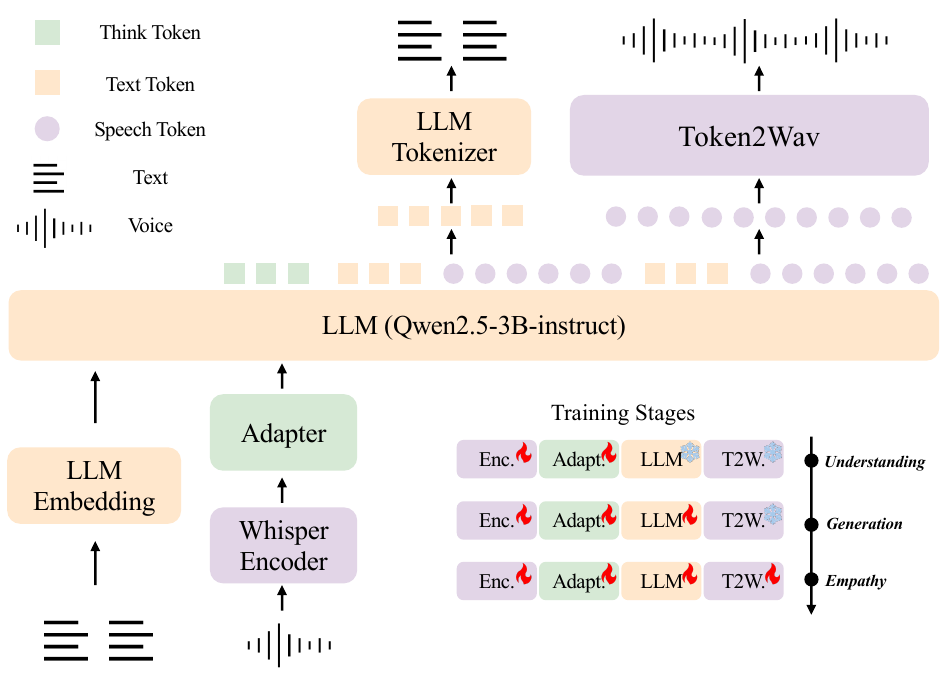}
    \caption{The architecture of the OSUM-EChat model. This end-to-end spoken dialogue model integrates various components, including a speech encoder, an adapter, an LLM decoder, and a token2wav module.}
    \label{fig:architecture}
\end{figure}


\section{OSUM-EChat}
\subsection{Architecture}
The overall architecture of OSUM-EChat consists of four key modules: a speech encoder, an adapter, an LLM decoder, and a token2wav module, as illustrated in Figure~\ref{fig:architecture}. The parameters of these modules are either learnable or frozen at different stages of training.

\textbf{Speech Encoder}
We use a pre-trained Whisper-Medium encoder~\cite{asr:whisper} as the speech encoder. The speech encoder transforms the 80-channel log-magnitude Mel spectrogram into high-dimensional feature vectors during input processing.

\textbf{Adapter}
The speech adapter consists of a three-layer 1D convolutional module, four transformer encoder layers~\cite{transformer}, and a final linear layer. The convolutional layer downsamples the input by a factor of four. The transformer encoder and final linear layer map the speech feature vectors from the speech encoder into the embedding space of the LLM.

\textbf{LLM}
The LLM decoder is based on Qwen2.5-3B-instruct~\cite{llm:qwen2.5}. To support the native multimodal paradigm, where the LLM handles speech modeling tasks, we extend the LLM's vocabulary by adding 4097 new tokens: 4096 correspond to speech tokens from the speech codebook of CosyVoice~\cite{tts:cosyvoice}, and one denotes the start or end of speech generation. The decoder receives speech representations from the speech adapter and generates text and speech tokens alternately, at 6:18. By leveraging the model's large parameter scale and inherent linguistic knowledge, this approach enables more fine-grained speech modeling.

\textbf{Token2wav}
The token2wav module is based on CosyVoice~\cite{tts:cosyvoice}, which includes a flow-matching model~\cite{flow-matching} for Mel spectrogram estimation and HiFi-GAN vocoder for audio reconstruction. The token2wav module converts the discrete speech tokens generated by the LLM into continuous audio waveforms, producing a 24 kHz, 16-bit PCM-format speech response that accurately reproduces and conveys paralinguistic information. The token2wav module is pre-trained from scratch to ensure smooth streaming and decoding.

\subsection{Training} 
Native multimodal models are well-suited for capturing paralinguistic nuances in speech~\cite{s2s:full_deeptalk}. However, previous implementations~\cite{s2s:full_baichuan_audio, s2s:full_glm4voice} have typically relied on large-scale industrial datasets to implicitly learn the complex associations between these nuances and conversational contexts. This dependency hinders further progress within the research community and limits their effectiveness in empathetic dialogue.
To enable OSUM-EChat to facilitate empathetic dialogue in resource-constrained environments, we propose a three-stage training strategy called \textbf{understanding-driven spoken dialogue}, which consists of three stages: understanding, generation, and empathy, as illustrated in Figure~\ref{fig:architecture}. To enhance empathetic capabilities, we explicitly decouple paralinguistic information using a \textbf{linguistic-paralinguistic dual think} mechanism during the empathy stage, which helps generate more empathetic responses.

\textbf{Stage 1: Understanding}
The goal of Stage 1 is to enable the LLM to comprehend both linguistic and paralinguistic information in speech. OSUM~\cite{osum} is an open-source speech understanding model designed to explore the potential of training speech language models with limited resources via an \textit{ASR+X} training strategy. We adopt OSUM's strategy with \textit{ASR+P}, where $P$ represents paralinguistic labels such as emotion, gender, age, and sound events.

In this stage, we jointly train multiple \textit{ASR+P} tasks, with only the encoder and adapter being trainable. After the model learns to recognize $P$, we expand the dataset by generating pseudo-labeled data that includes all $P$ labels. This enables the model to recognize all $P$ labels simultaneously, providing a foundation for integration into dialogue tasks.
The loss function for this stage is expressed as:
\begin{equation}
L_{\textit{S1}} = \sum_{i=1}^{N} L_{\textit{ASR}}(x_i, \hat{x_i}) + \sum_{j=1}^{M} L_{\textit{P}}(p_{j}, \hat{p}_{j}),
\end{equation}
where $x_i$ and $\hat{x}i$ are the ground-truth and predicted text tokens for the ASR task, and $p{j}$ and $\hat{p}_{j}$ correspond to the $j$-th paralinguistic label. N and M represent the number of ASR and $P$ tokens, respectively.
This phase establishes the essential groundwork for generating empathetic responses by consolidating the understanding capabilities required for further paralinguistic modeling.

\begin{figure}[t]
    \centering
    \includegraphics[width=\linewidth]{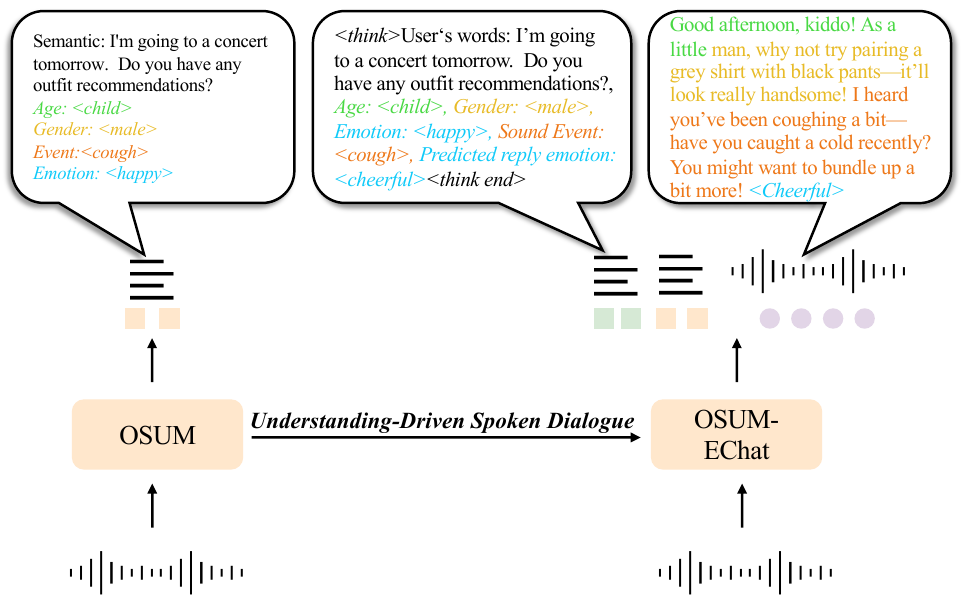}
    \caption{Understanding-driven spoken dialogue training strategy and linguistic-paralinguistic dual think mechanism.}
    \label{fig:dualthink}
\end{figure}

\textbf{Stage 2: Generation}
In this stage, the aim is to equip the OSUM-based understanding model with speech generation capabilities. We employ a two-step training process: text-to-speech (TTS) generation and speech-to-speech (S2S) generation. We also add text-to-text (T2T) data to retain the model's intelligence.

Text-to-Speech (TTS):
We fine-tune the LLM on TTS task data in the TTS phase to generate speech tokens. The model's input is text, and the output is speech tokens, with the LLM being the only learnable component.
The loss function for this phase is:
\begin{equation}
L_{\textit{S2-TTS}} = \sum_{i=1}^{S} L_{\textit{speech}}(y_i, \hat{y}_i),
\end{equation}
where $y_i$ and $\hat{y}_i$ represent the ground-truth and predicted speech tokens.

Speech-to-Speech (S2S):
Building on the TTS phase, we further perform joint training on speech-to-speech dialogue data. In this phase, speech encoder, adapter, and LLM are all learnable components. During training, the LLM receives the input speech representation from the adapter and generates both text and speech tokens.
The S2S task is divided into non-streaming and streaming modes. In non-streaming mode, the model first outputs text tokens, followed by speech tokens, while in streaming mode, it alternates between text and speech tokens, maintaining a 6:18 ratio for stable speech generation.
The loss function for the S2S phase is:
\begin{equation}
L_{\textit{S2-S2S}} = \sum_{i=1}^{N} L_{\textit{text}}(x_i, \hat{x}_i) + \sum_{j=1}^{S} L_{\textit{speech}}(y_j, \hat{y}_j),
\end{equation}
where $x_i$ and $\hat{x}_i$ are the ground-truth and predicted text tokens, and $y_j$ and $\hat{y}_j$ are the corresponding speech tokens. This stage provides the necessary groundwork for generating empathetic responses by equipping the model with the ability to produce spoken dialogue.

\textbf{Stage 3: Empathy}
In this stage, linguistic and paralinguistic information from speech understanding is integrated into the dialogue generation process, significantly improving the model's ability to produce contextually coherent and empathetic responses. This stage bridges the gap between understanding and dialogue generation through a \textit{linguistic-paralinguistic dual think} mechanism.

\textbf{Linguistic-Paralinguistic Dual Think}
We introduce a dedicated Chain of Thought (CoT) process prior to the model generating text and speech responses. As shown in Figure~\ref{fig:dualthink}, the output of the CoT phase is enclosed between the $\langle think \rangle$ and $\langle think\ end\rangle$ tokens, ensuring clear separation from the subsequent response generation. During this phase, the model first identifies the linguistic information in the user's speech, followed by inferring paralinguistic details.
The corresponding loss function is:
\begin{equation}
L_{\textit{S3}} = \sum_{i=1}^{N} L_{\textit{dual-think}}(t_i, \hat{t}_i) + L_{\textit{S2-S2S}},
\end{equation}
where $t_i$, $\hat{t}_i$ represent the ground-truth and predicted think tokens from the CoT phase.
After completing the CoT phase, the model integrates linguistic and paralinguistic insights to generate the appropriate text and speech responses, enabling it to function as a fully integrated spoken dialogue system capable of empathetic communication.

\section{EChat-200K \&  EChat-eval}
\subsection{Data Pipline}
To construct empathetic conversational data, we design a comprehensive empathy data generation pipeline. First, we use DeepSeek~\cite{llm:deepseekv2} with customized prompts to generate questions for empathy-related scenarios, where each question is annotated with paralinguistic information. Second, we employ CosyVoice2~\cite{tts:cosyvoice2}, a voice-controllable text-to-speech (TTS) system, to generate input query speech. This generated speech is then concatenated with supplementary audio clips containing relevant sound events to form a richer, more contextually immersive input. Finally, DeepSeek is reused to generate a detailed empathetic response based on the generated query text and corresponding paralinguistic labels; this response is further converted into emotionally resonant speech using CosyVoice2. The entire process is detailed in Appendix A.1 of the supplementary material.

\begin{figure}[h]
    \centering
    \includegraphics[width=\linewidth]{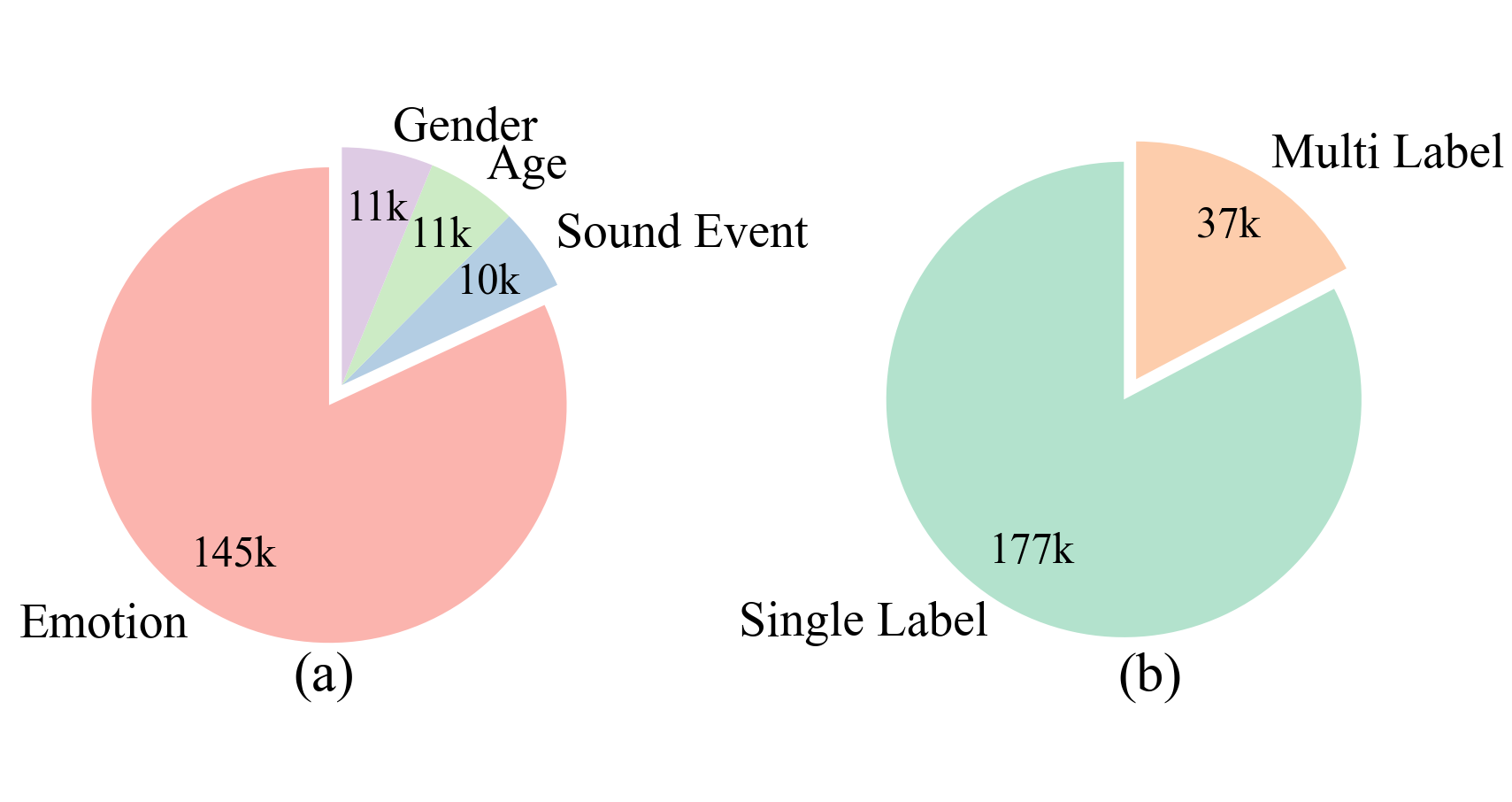}
    \caption{The distribution of paralinguistic labels in the \textit{EChat-200K}. (a) The distribution of single-label data. (b) Comparison between single-label and multi-label data.}
    \label{fig: EChat data}
\end{figure}
\subsection{EChat-200K Dataset}
To enable OSUM-EChat to process paralinguistic cues crucial for empathetic conversations, we construct the \textit{EChat-200K} dataset, consisting of approximately 200k conversations, following the data pipeline outlined previously.

As shown in Figure~\ref{fig: EChat data}, the \textit{EChat-200K} dataset is divided into two categories: single-label empathetic data and multi-label empathetic data. The single-label data focuses on one paralinguistic cue per interaction, simplifying the learning process by narrowing the model's focus. We create single-label data for emotion, gender, age, and sound events, with emotion being the most critical cue for empathy. Thus, we generate a larger portion of the data for emotion.
In contrast, the multi-label data encourages the model to integrate multiple paralinguistic cues, generating more comprehensive and contextually accurate responses. A key observation during data generation is that the emotional expressiveness of synthesized audio from emotional TTS systems often lags behind that of real human speech. To mitigate the risk of overfitting to synthetic audio, we incorporate a proportion of both single-label and multi-label data with real audio input queries, including 107k real labels in the single-label emotion data and 10k real labels in the multi-label data. The corresponding response audio is then generated through the latter part of the data pipeline.

\subsection{EChat-eval Bench}
To systematically evaluate a model’s capacity to capture paralinguistic cues critical for empathetic dialogue, we propose the \textit{EChat-eval} benchmark, which aligns with the task taxonomy of the \textit{EChat-200K} training corpus and is constructed using the aforementioned data generation pipeline. Each task within this benchmark contains 280 entries. To mitigate the discrepancy between synthesized data and real human dialogue scenarios, approximately one-third of these entries are derived from real recordings. 

For fair and consistent performance assessment, we utilize a standardized evaluation pipeline. ChatGPT-4o \cite{gpt4o} acts as the core scoring model, followed by VoiceBench~\cite{voicebench}, and emotion2vec-Large \cite{emotion2vec} is deployed to extract emotional labels from response speech. Concretely, the query text, its associated paralinguistic labels, the response text, and the emotional labels derived from the response speech are jointly input into ChatGPT-4o. It then generates scores across three dimensions, and the average of these scores is computed as the final evaluation result. For the sake of transparency, examples of the benchmark and the prompts used in automated scoring are included in Appendix A.2 of the supplementary material.

\begin{table*}[h]
\centering
\resizebox{1.0\linewidth}{!}{
\begin{tabular}{@{}
>{\raggedright\arraybackslash}m{0.2\linewidth}
>{\centering\arraybackslash}m{0.1\linewidth}
>{\centering\arraybackslash}m{0.44\linewidth}
>{\centering\arraybackslash}m{0.14\linewidth}
>{\centering\arraybackslash}m{0.07\linewidth}
>{\centering\arraybackslash}m{0.07\linewidth}
>{\centering\arraybackslash}m{0.07\linewidth}@{}}
\toprule
\textbf{Task} & \textbf{Type} & \textbf{Data and Details} & \textbf{Volume} & \textbf{Stage 1} & \textbf{Stage 2} & \textbf{Stage 3} \\
\midrule
\textbf{Speech Understanding} & ASR+P &
Data of OSUM~\cite{osum} &
50 Kh & \checkmark & \checkmark & \checkmark  \\
\midrule

\textbf{Text Q\&A} & T2T &
Data of Alpaca-CoT~\cite{all_t2t} &
40 Mp & - & \checkmark & \checkmark \\
\midrule

\multirow{3}{*}{\textbf{Speech Synthesis}} & \multirow{3}{*}{TTS} &
Emilia~\cite{emilia} & 100 Kh & \multirow{3}{*}{-} & \multirow{3}{*}{\checkmark} & \multirow{3}{*}{\checkmark} \\
& & LibriHeavy~\cite{libriheavy} & 10 Kh & & & \\
& & Internal high-quality data & 2 Kh & & & \\
\midrule

\multirow{3}{=}{\textbf{Audio Q\&A}} & \multirow{3}{*}{S2S} &
Selected ASR audio data as query, LLM generated and TTS synthesized response & 15 Kh & \multirow{3}{*}{-} & \multirow{3}{*}{\checkmark} & \multirow{3}{*}{\checkmark} \\
& &Linguistic dialogue based text data& 4 Kh & & & \\
& &  Internal Multi-speaker podcast dialogue& 1.5 Kh & & & \\
\midrule

\textbf{Empathetic dialogue} & S2S &
EChat-200K corpus &
0.2 Kh & - & - & \checkmark \\
\bottomrule
\end{tabular}
}
\caption{Data overview for OSUM-EChat tasks. Here, `Kh' refers to one thousand hours; `Mp' refers to one million pairs.   }
\label{tab:data_overview}
\end{table*}

\section{Experiments}
\subsection{Experimental Setup}
\textbf{Dataset}
To ensure the completeness and effectiveness of OSUM-EChat in resource-limited empathetic dialogue scenarios, we construct datasets covering four core tasks: speech understanding, text Q\&A, speech synthesis, Audio Q\&A, and empathetic dialogue. Details of the dataset are provided in Table~\ref{tab:data_overview}, including the data duration and the specific training phases involved. More comprehensive information on its contents can be found in Appendix A.3.

\textbf{Training Details}
We employ the WeNet framework~\cite{wenet} for training, using the AdamW optimizer~\cite{adamw} with the following configuration: betas = (0.9, 0.999), weight decay = 0.01, and an initial learning rate of 5e-6. To stabilize training and accelerate convergence, we apply a warmup strategy~\cite{transformer} with 8,000 steps, gradually increasing the learning rate to avoid abrupt fluctuations in the early stages. To prevent gradient exploding, we use gradient clipping with a threshold of 5, and apply gradient accumulation over four mini-batches to balance memory efficiency and computational performance. For large-scale training under memory constraints, we adopt a dynamic batch size strategy, limiting the total token count per batch for LLM inputs to 2,800 tokens. This approach reduces memory consumption while optimizing computational efficiency. Training is conducted on 64 ShengTeng 910B GPUs. Each training run starts with a warmup from step 0, and all other configurations follow the defaults outlined above unless specified otherwise.
The entire process is detailed in Appendix B.

\textbf{Contrastive Models}
We use current end-to-end spoken dialogue models and empathic models for comparison, including Qwen2.5-Omni~\cite{s2s:align_qwen2.5_omni}, Freeze-Omni~\cite{s2s:align_freeze_omni}, GLM4-Voice~\cite{s2s:full_glm4voice}, Baichuan-Audio~\cite{s2s:full_baichuan_audio}, Kimi-Audio~\cite{kimi-audio}, Moshi~\cite{s2s:full_moshi}, DeepTalk~\cite{s2s:full_deeptalk}, and MinMo~\cite{s2s:align_minmo}, et al.

\begin{table}[t]
\centering
\resizebox{\linewidth}{!}{
\begin{tabular}{lccccc}
\toprule
\textbf{Model} & \textbf{Emotion} & \textbf{Age} & \textbf{Gender} & \textbf{Event} & \textbf{Multi} \\ \midrule
Qwen2.5-Omni & 44.8 & 52.6 & 43.3 & -  & 63.7 \\
Freeze-Omni & 44.0 & 31.7 & 47.3 & -  & 49.6 \\
GLM-4-Voice & 48.7 & 44.7 & 42.4 & - & 60.3 \\
Baichuan-Audio & 41.5 & 28.6 & 41.6 & - & 45.6 \\
Kimi-Audio & 45.8 & 34.2 & 36.0 &- & 57.4 \\
OpenS2S & 46.6 & 62.1 & 37.9 & - & 57.3 \\
\midrule
\textbf{OSUM-EChat} & \textbf{58.0} &\textbf{ 63.1} & \textbf{62.7} & \textbf{87.1} & \textbf{72.0} \\
\ \ \ - U-Driven & 53.4 & 35.4 & 41.0 & 52.9 & 63.1 \\
\ \ \ - Dual Think & 54.5 & 42.3 & 61.3 & 61.1 & 56.4 \\
\ \ \ \ \ \ - U-Driven & 51.6 & 40.3 & 39.1 & 10.4 & 54.2 \\
\bottomrule
\end{tabular}
}
\caption{Automatic evaluation results on \textit{EChat-eval} benchmark.
Here, `U-Driven' refers to the \textit{understanding-driven spoken dialogue} training strategy, and `Dual Think' refers to the \textit{linguistic-paralinguistic dual think} mechanism.
}
\label{tab:empathic_comparison}
\end{table}

\subsection{Main Results}
\textbf{Results of Empathetic Intelligence}
We evaluate OSUM-EChat's capabilities in empathic spoken conversation using the \textit{EChat-eval} benchmark, which includes diverse test cases across multiple paralinguistic dimensions. The evaluation results are presented in Table~\ref{tab:empathic_comparison}. OSUM-EChat demonstrates strong performance, achieving an excellent GPT-4 score across various empathic conversation tasks, including both single-label and multi-label scenarios. Notably, OSUM-EChat attains a GPT-4 score of 72.0 in the multi-label scenario, outperforming other end-to-end spoken conversation models.
This suggests that OSUM-EChat excels at simultaneously recognizing multiple paralinguistic cues and generating comprehensive responses. We further observe that scores in multi-label scenarios are generally higher, as the scoring model does not strictly require responses to incorporate all paralinguistic information. Instead, it tends to assign higher scores when even partial paralinguistic cues are addressed.
Additionally, mainstream end-to-end spoken conversation models have difficulty perceiving diverse acoustic events in input speech and generating appropriate responses. In contrast, OSUM-EChat achieves a high score of 87.1, underscoring its superior empathic dialogue capabilities despite being trained on relatively limited data. This result confirms its outperformance over both previous spoken dialogue models and contemporary empathic dialogue models~\cite{s2s:align_qwen2.5_omni, wang2025opens2s}.
We also find that emotion2vec-large~\cite{emotion2vec} shows relatively limited emotional diversity in the speech outputs of other end-to-end models, with neutral and happy tones predominating. This limitation constrains their scores, leading to relatively similar performance metrics and indicating a lack of emotional richness in their speech outputs.

To validate the effectiveness of the \textit{EChat-eval} benchmark, we conduct human evaluations on OSUM-EChat, Qwen2.5-Omni~\cite{s2s:align_qwen2.5_omni}, and the commercial spoken dialogue model Doubao~\footnote{\url{https://www.volcengine.com/docs/6561/1594360}}. As shown in Table~\ref{tab:human_empathic}, OSUM-EChat outperforms Qwen2.5-Omni~\cite{s2s:align_qwen2.5_omni}, but still lags behind commercial counterparts on empathetic dialogue test cases within the emotional dimension. For other paralinguistic dimensions, we find that commercial systems still cannot perceive these paralinguistic cues.
Notably, our analysis reveals differences between human scores and automated ratings. This divergence primarily stems from labeling errors in emotion2vec-large~\cite{emotion2vec} and hallucinatory outputs from ChatGPT-4o~\cite{gpt4o}, suggesting room for refinement in our automated evaluation pipeline. However, consistent model rankings between human and automated evaluations further confirm the validity of our automated scoring framework.

\textbf{Ablation Studies}
To validate the effectiveness of our proposed training strategy and the \textit{linguistic-paralinguistic dual think} mechanism, we conduct a series of ablation experiments. As shown in the lower half of Table~\ref{tab:empathic_comparison}, the first experiment \textit{U-Driven} remove the \textit{understanding-driven spoken dialogue} training strategy from the full OSUM-EChat model, relying solely on ASR data for modality alignment while maintaining the linguistic-paralinguistic dual think mechanism. The second experiment \textit{Dual Think} remove the \textit{dual think} mechanism, specifically omitting the linguistic-paralinguistic dual think mechanism. The third experiment remove both strategies.
The results align with our expectations, demonstrating that utilizing a speech understanding model for spoken conversation tasks and incorporating the linguistic-paralinguistic dual think mechanism together significantly enhance empathic conversation capabilities.

\begin{table}[t]
\centering
\resizebox{\linewidth}{!}{
\begin{tabular}{lccccc}
\toprule
\textbf{Model} & \textbf{Emotion} & \textbf{Age} & \textbf{Gender} & \textbf{Event} & \textbf{Multi} \\ \midrule
Doubao$^\dagger$ & \textbf{78.0 }& 42.2 & 39.0 & - & 60.7 \\
Qwen2.5-Omni & 54.9 & 55.0 & 49.2 & - & 67.3 \\
OSUM-EChat& 72.0 &\textbf{ 68.9 }& \textbf{65.8 }& \textbf{88.0} & \textbf{73.3} \\
\bottomrule
\end{tabular}
} 

\caption{Human evaluation results of representative models on the \textit{EChat-eval} benchmark. \small $^\dagger$ ByteDance's commercial system with response from a single fixed speaker. 
}
\label{tab:human_empathic}
\end{table}

\begin{table}[t]
\centering
\resizebox{\linewidth}{!}{
\begin{tabular}{@{}lcccccc@{}}
\toprule
\textbf{Model} & \multicolumn{2}{c}{\textbf{LLaMA Question}} & \multicolumn{2}{c}{\textbf{TriviaQA}} & \multicolumn{2}{c}{\textbf{Web Question}} \\ \cmidrule(lr){2-3} \cmidrule(lr){4-5} \cmidrule(lr){6-7}
& \textbf{s2t} & \textbf{s2s} & \textbf{s2t} & \textbf{s2s} & \textbf{s2t} & \textbf{s2s}  \\ \midrule
ChatGPT-4o & - & \textbf{71.7} & - & \textbf{69.7 }& - & \textbf{51.6}   \\
GLM-4-Voice & 64.7 & 50.7 & 39.1 & 26.5 & 32.2 & 15.9 \\
Moshi & 62.3 & 21.0 & 22.8 & 7.30 & 26.2 & 9.20 \\
DeepTalk & 65.0 & 59.7 &\textbf{ 41.2 }& 27.5 & 34.4 & 23.1 \\
MinMo & \textbf{78.9} & 64.1 & 48.3 & 37.5 &\textbf{ 55.0 }& 39.9  \\
OSUM-EChat & 55.3 & 57.0 & 33.7 & 31.7 & 30.4 & 28.6 \\
\bottomrule
\end{tabular}
}
\caption{Performance on spoken question answering.}
\label{tab:qa_sanjianke}
\end{table}



\subsection{Basic Results}
\textbf{Results of Linguistic Intelligence}
We evaluate the dialogue capabilities of OSUM-EChat using the UltraEval-Audio benchmark\footnote{https://github.com/OpenBMB/UltraEval-Audio} and Voicebench~\cite{voicebench}. The evaluation covers both speech-to-text (S2T) and speech-to-speech (S2S) dialogue modes, with text-to-text (T2T) dialogue results provided in Appendix C. Results from UltraEval-Audio are presented in Table~\ref{tab:qa_sanjianke}, and Voicebench results are detailed in Appendix C. In the S2S dialogue mode, OSUM-EChat performs less effectively than commercial models (\textit{ChatGPT-4o~\cite{gpt4o}}) and modularly aligned multimodal models (\textit{MinMo~\cite{s2s:align_minmo}}), but generally outperforms or achieves comparable performance to other native multimodal models. In the S2T dialogue mode, OSUM-EChat produces results similar to those of other native multimodal models. These findings demonstrate that OSUM-EChat exhibits good linguistic capabilities.

\begin{table}[t]
\centering
\footnotesize
\resizebox{0.7\linewidth}{!}{
\begin{tabular}{lccc}
\toprule
\textbf{Model} & \textbf{test-zh} & \textbf{test-en} \\ \midrule
Seed-TTS & \textbf{1.12} &\textbf{ 2.25}  \\
CosyVoice & 3.63 & 4.29 \\
Qwen2.5-Omni & 1.70 & 2.72 \\
GLM-4-Voice & 2.10 & 2.91  \\
 OSUM-EChat & 3.00 &3.99\\  \bottomrule
\end{tabular}
}
\caption{WER\%($\downarrow$) and CER\%($\downarrow$) results of OSUM-EChat and recent spoken dialogue models on the SEED test sets.}
\label{tab:tts_metrics}
\end{table}

\textbf{Results of Speech Understanding}
We first evaluate OSUM-EChat against state-of-the-art speech understanding models across five tasks—ASR, emotion recognition, gender recognition, age recognition, and sound event recognition—using open-source test sets. As shown in Table~\ref{tab:um_results}, OSUM-EChat-3B performs comparably to OSUM-7B~\cite{osum} and Qwen2-audio-7B~\cite{um:qwen2audio}, despite having 60\% fewer parameters.

To assess the effectiveness of the \textit{dual think} mechanism in knowledge transfer, we evaluate the accuracy of paralinguistic label in the think token of CoT phase using public test sets for speech understanding tasks. \textit{S2S\_think\_token}'s performance in ASR and paralinguistic information is comparable to \textit{ASR+P}, confirming that the mechanism can effectively transfer speech understanding knowledge to spoken dialogue tasks.
Furthermore, to verify that these performance gains stemmed from knowledge transfer from the comprehension task rather than from the \textit{EChat-200K} dataset itself, we use \textit{U-Driven} ablation to evaluate the accuracy of identifying relevant paralinguistic cues during the think token of CoT phase. The results show that this configuration performed significantly worse than the full model.
Additional results from public speech understanding test sets can be found in Appendix C.

\textbf{Results of Speech Synthesis}
We evaluate the speech synthesis performance of OSUM-EChat using the SEED test sets~\cite{seed-tts}, with the results presented in Table~\ref{tab:tts_metrics}. These results confirm that OSUM-EChat demonstrates functional TTS capabilities. We find that OSUM-EChat's TTS performance marginally outperforms the CosyVoice model; however, there is still a noticeable gap when compared to industry-leading TTS models and other end-to-end spoken dialogue models~\cite{seed-tts, s2s:align_qwen2.5_omni, s2s:full_glm4voice}. 

Additional results from paralinguistic instruction-following tests, along with other evaluations, are provided in Appendix C.



\begin{table}[t]
\centering
\footnotesize
\resizebox{\linewidth}{!}{
\begin{tabular}{lccccc}
\toprule
\textbf{Model} & \textbf{ASR} & \textbf{Age} & \textbf{Gender} & \textbf{Emotion} & \textbf{Sound} \\
 & \textbf{AL-2} & \textbf{Kaggle} & \textbf{Kaggle} & \textbf{MER23} & \textbf{VS} \\ 
\midrule
Qwen2audio & 3.01 & 35.5 & 97.3 & - & \textbf{93.3 }\\
OSUM & 2.81 & 76.5 & 99.4 & \textbf{86.4} & 82.6 \\ 
\midrule
OSUM-EChat &  &  &  &  &  \\ 
\ \ ASR+P & 2.94 & 83.8 &\textbf{ 99.7} & 85.9 & 82.6 \\
\ \ S2S\_{think\_token} & \textbf{2.80} & \textbf{84.1}& 98.6 & 79.0 & 80.2 \\
\ \ \ \ - U-Driven & 2.83 & 84.0 & 74.3& 25.1 &68.2 \\
\bottomrule
\end{tabular}
}
\caption{Performance comparison of different models on speech understanding tasks. Here, `U-Driven' refers to the \textit{understanding-driven spoken dialogue} training strategy; `AL-2' denotes the AISHELL-2 test set; and `VS' refers to the VocalSound test set.}
\label{tab:um_results}
\end{table}

\section{Related Work}

\subsection{End-to-End Spoken Dialogue System}
Speech interaction is a pivotal human-computer interaction mode. End-to-end solutions using unified models for speech and text processing have become dominant, falling into two categories~\cite{s2s:align_minmo}: modular-aligned and native multimodal models. Modular-aligned models (e.g., Qwen2.5-Omni~\cite{s2s:align_qwen2.5_omni}, Minmo~\cite{s2s:align_minmo}, LLaMA-Omni~\cite{s2s:align_llmama_omni}, Freeze-Omni~\cite{s2s:align_freeze_omni}) connect LLMs to audio encoders/decoders via adapters, preserving LLM language capabilities but relying on audio decoders for paralinguistic information—underutilizing LLM capabilities and posing deployment efficiency issues~\cite{s2s:full_deeptalk}.
In contrast, native speech-language models (e.g., Mini-Omni~\cite{MiniCPM-o}, Moshi~\cite{s2s:full_moshi}, GLM-4-Voice~\cite{llm:chatglm}, mini-Omni~\cite{s2s:full_mini_omni}, LUCY~\cite{s2s:full_lucy}) directly integrate LLMs for text and speech decoding, reducing latency, simplifying deployment, and enabling richer paralinguistic modeling. Yet they require large-scale paired speech data; without it, their paralinguistic ability remains limited and data-dependent. OSUM-EChat mitigates these limitations via an \textit{understanding-driven spoken dialogue} strategy, transferring pre-trained speech understanding capabilities to dialogue tasks to reduce reliance on large dialogue datasets.

\subsection{Empathetic Dialogue}
Significant progress has been made in empathetic dialogue systems. Early research focuses on text-based dialogue~\cite{t2t_empathetic_1, t2t_empathetic_2}: ELIZA~\cite{ELIZA_first_e_chat} simulated empathy via simple pattern matching, while SoulChat~\cite{t2t_empathetic_SoulChat} fine-tuned LLMs for multi-turn empathetic dialogues, notably in psychological counseling. In the S2T domain, BLSP-Emo~\cite{s2t_empathetic} and E-chat~\cite{e-chat} advanced emotional speech recognition, enhancing spoken dialogue systems’ emotional sensitivity.
Recently, SpeechGPT-Gen~\cite{speechGPT-gen} applied a cascaded framework in spoken dialogue, refining responses through multiple stages. Commercial systems like ChatGPT-4o~\cite{gpt4o} and Doubao set benchmarks for end-to-end multimodal empathetic conversations. Among open-source systems: Modular-aligned models such as OpenS2S~\cite{wang2025opens2s} achieved emotion-centric capabilities via speech-to-speech datasets; Goal-SLM~\cite{goat-slm} improves paralinguistic perception through a bimodal head but remains limited to single-cue perception and weak generation. Native multimodal models like Step-Audio2~\cite{step-audio2} showed potential in coordinating emotion perception and response generation but only possessed paralinguistic perception capabilities. OSUM-EChat, adopting a native multimodal architecture, enables simultaneous perception of multiple paralinguistic cues and generation of corresponding responses.

\section{Limitation}

While this study makes a modest contribution to advancing empathetic dialogue research, it has several limitations. First, dynamic paralinguistic scenarios—such as transitions in emotion (e.g., from sadness to joy) or multi-speaker interactions involving diverse genders and ages—remain underexplored. Second, similar to others, EChat-eval’s automatic scoring system faces several challenges, including inaccurate emotional label extraction and difficulty in scoring LLM hallucinations, which can differ substantially from human evaluations.
Future work should aim to explore complex conversational scenarios and refine the automated scoring system.


\section{Conclusion}
In this work, we present OSUM-EChat, an end-to-end empathetic spoken dialogue system designed to enhance paralinguistic modeling under resource constraints. By leveraging an \textit{understanding-driven spoken dialogue} training strategy and a \textit{linguistic-paralinguistic dual thinking} mechanism, OSUM-EChat improves empathetic response generation while reducing reliance on large-scale empathetic spoken dialogue datasets. Complementing this, we introduce the \textit{EChat-200K} dataset and \textit{EChat-eval} benchmark, which offer critical resources for training and evaluating empathetic dialogue systems. Experimental results validate that OSUM-EChat outperforms existing spoken dialogue models in empathy, confirming its effectiveness. Finally, we contribute a complete open-source ecosystem to facilitate future research in empathetic dialogue.

\bibliography{aaai2026}

\begin{thebibliography}{62}
\providecommand{\natexlab}[1]{#1}

\bibitem[{{AISHELL Tech Co Ltd}(2024)}]{aishell_page}
{AISHELL Tech Co Ltd}. 2024.
\newblock {Data Products}.
\newblock \url{https://www.aishelltech.com/General_Datasets}, Last accessed on 2025-1-9.

\bibitem[{Anastassiou et~al.(2024)Anastassiou, Chen, Chen, Chen, Chen, Chen, and et~al.}]{seed-tts}
Anastassiou, P.; Chen, J.; Chen, J.; Chen, Y.; Chen, Z.; Chen, Z.; and et~al. 2024.
\newblock Seed-TTS: {A} Family of High-Quality Versatile Speech Generation Models.
\newblock \emph{CoRR}, abs/2406.02430.

\bibitem[{Bu et~al.(2017)Bu, Du, Na, Wu, and Zheng}]{aishell1}
Bu, H.; Du, J.; Na, X.; Wu, B.; and Zheng, H. 2017.
\newblock {AISHELL-1}: {An} open-source Mandarin speech corpus and a speech recognition baseline.
\newblock In \emph{Proceedings of the Conference of the Oriental Chapter of the International Coordinating Committee on Speech Databases and Speech {I/O} Systems and Assessment (O-COCOSDA)}.

\bibitem[{Busso et~al.(2008)Busso, Bulut, Lee, Kazemzadeh, Mower, Kim, Chang, Lee, and Narayanan}]{busso2008iemocap}
Busso, C.; Bulut, M.; Lee, C.-C.; Kazemzadeh, A.; Mower, E.; Kim, S.; Chang, J.~N.; Lee, S.; and Narayanan, S.~S. 2008.
\newblock {IEMOCAP}: {Interactive} emotional dyadic motion capture database.
\newblock \emph{Language Resources and Evaluation (LREC)}, 42: 335--359.

\bibitem[{Chen et~al.(2021)Chen, Chai, Wang, Du, Zhang, Weng, Su, Povey, Trmal, Zhang, Jin, Khudanpur, Watanabe, Zhao, Zou, Li, Yao, Wang, You, and Yan}]{gigaspeech}
Chen, G.; Chai, S.; Wang, G.; Du, J.; Zhang, W.; Weng, C.; Su, D.; Povey, D.; Trmal, J.; Zhang, J.; Jin, M.; Khudanpur, S.; Watanabe, S.; Zhao, S.; Zou, W.; Li, X.; Yao, X.; Wang, Y.; You, Z.; and Yan, Z. 2021.
\newblock GigaSpeech: An Evolving, Multi-Domain {ASR} Corpus with 10, 000 Hours of Transcribed Audio.
\newblock 3670--3674. {ISCA}.

\bibitem[{Chen et~al.(2025{\natexlab{a}})Chen, Li, Song, Deng, Yao, Zhang, Lv, Zhu, Kang, Lian et~al.}]{goat-slm}
Chen, H.; Li, Z.; Song, Y.; Deng, W.; Yao, Y.; Zhang, Y.; Lv, H.; Zhu, X.; Kang, J.; Lian, J.; et~al. 2025{\natexlab{a}}.
\newblock GOAT-SLM: A Spoken Language Model with Paralinguistic and Speaker Characteristic Awareness.
\newblock \emph{arXiv preprint arXiv:2507.18119}.

\bibitem[{Chen et~al.(2025{\natexlab{b}})Chen, Chen, Chen, Chen, Chen, Deng, Du, Gao, Gao, Gao, Li, Lv, Liu, Luo, Ma, Ni, Shi, Tang, Wang, Wang, Wang, Wang, Xu, Yu, Yan, Yang, Yang, Yang, Yang, Zhao, Zhang, Zhang, Zhao, Zhang, Zhang, and Zhou}]{s2s:align_minmo}
Chen, Q.; Chen, Y.; Chen, Y.; Chen, M.; Chen, Y.; Deng, C.; Du, Z.; Gao, R.; Gao, C.; Gao, Z.; Li, Y.; Lv, X.; Liu, J.; Luo, H.; Ma, B.; Ni, C.; Shi, X.; Tang, J.; Wang, H.; Wang, H.; Wang, W.; Wang, Y.; Xu, Y.; Yu, F.; Yan, Z.; Yang, Y.; Yang, B.; Yang, X.; Yang, G.; Zhao, T.; Zhang, Q.; Zhang, S.; Zhao, N.; Zhang, P.; Zhang, C.; and Zhou, J. 2025{\natexlab{b}}.
\newblock MinMo: {A} Multimodal Large Language Model for Seamless Voice Interaction.
\newblock \emph{CoRR}, abs/2501.06282.

\bibitem[{Chen et~al.(2023)Chen, Xing, Lin, Zheng, Wang, Liu, and Xu}]{t2t_empathetic_SoulChat}
Chen, Y.; Xing, X.; Lin, J.; Zheng, H.; Wang, Z.; Liu, Q.; and Xu, X. 2023.
\newblock SoulChat: Improving LLMs' Empathy, Listening, and Comfort Abilities through Fine-tuning with Multi-turn Empathy Conversations.
\newblock In Bouamor, H.; Pino, J.; and Bali, K., eds., \emph{{EMNLP} 2023}, 1170--1183. Association for Computational Linguistics.

\bibitem[{Chen et~al.(2024)Chen, Yue, Zhang, Gao, Tan, and Li}]{voicebench}
Chen, Y.; Yue, X.; Zhang, C.; Gao, X.; Tan, R.~T.; and Li, H. 2024.
\newblock VoiceBench: Benchmarking LLM-Based Voice Assistants.
\newblock \emph{CoRR}, abs/2410.17196.

\bibitem[{Chu et~al.(2024)Chu, Xu, Yang, Wei, Wei, Guo, Leng, Lv, He, Lin, Zhou, and Zhou}]{um:qwen2audio}
Chu, Y.; Xu, J.; Yang, Q.; Wei, H.; Wei, X.; Guo, Z.; Leng, Y.; Lv, Y.; He, J.; Lin, J.; Zhou, C.; and Zhou, J. 2024.
\newblock Qwen2-Audio Technical Report.
\newblock \emph{CoRR}, abs/2407.10759.

\bibitem[{{Datatang Tech Co Ltd}(2024)}]{datatang_page}
{Datatang Tech Co Ltd}. 2024.
\newblock Data Products.
\newblock \url{https://www.datatang.com/speechRecognition}, Last accessed on 2025-1-9.

\bibitem[{D{\'{e}}fossez et~al.(2024)D{\'{e}}fossez, Mazar{\'{e}}, Orsini, Royer, P{\'{e}}rez, J{\'{e}}gou, Grave, and Zeghidour}]{s2s:full_moshi}
D{\'{e}}fossez, A.; Mazar{\'{e}}, L.; Orsini, M.; Royer, A.; P{\'{e}}rez, P.; J{\'{e}}gou, H.; Grave, E.; and Zeghidour, N. 2024.
\newblock Moshi: a speech-text foundation model for real-time dialogue.
\newblock \emph{CoRR}, abs/2410.00037.

\bibitem[{Du et~al.(2018)Du, Na, Liu, and Bu}]{aishell2}
Du, J.; Na, X.; Liu, X.; and Bu, H. 2018.
\newblock {AISHELL-2}: {Transforming Mandarin ASR} research into industrial scale.
\newblock \emph{arXiv preprint arXiv:1808.10583}.

\bibitem[{Du et~al.(2024{\natexlab{a}})Du, Chen, Zhang, Hu, Lu, Yang, Hu, Zheng, Gu, Ma, Gao, and Yan}]{tts:cosyvoice}
Du, Z.; Chen, Q.; Zhang, S.; Hu, K.; Lu, H.; Yang, Y.; Hu, H.; Zheng, S.; Gu, Y.; Ma, Z.; Gao, Z.; and Yan, Z. 2024{\natexlab{a}}.
\newblock CosyVoice: {A} Scalable Multilingual Zero-shot Text-to-speech Synthesizer based on Supervised Semantic Tokens.
\newblock \emph{CoRR}, abs/2407.05407.

\bibitem[{Du et~al.(2024{\natexlab{b}})Du, Wang, Chen, Shi, Lv, Zhao, Gao, Yang, Gao, Wang et~al.}]{tts:cosyvoice2}
Du, Z.; Wang, Y.; Chen, Q.; Shi, X.; Lv, X.; Zhao, T.; Gao, Z.; Yang, Y.; Gao, C.; Wang, H.; et~al. 2024{\natexlab{b}}.
\newblock Cosyvoice 2: Scalable streaming speech synthesis with large language models.
\newblock \emph{arXiv preprint arXiv:2412.10117}.

\bibitem[{Fang et~al.(2025)Fang, Guo, Zhou, Ma, Zhang, and Feng}]{s2s:align_llmama_omni}
Fang, Q.; Guo, S.; Zhou, Y.; Ma, Z.; Zhang, S.; and Feng, Y. 2025.
\newblock LLaMA-Omni: Seamless Speech Interaction with Large Language Models.
\newblock In \emph{The Thirteenth International Conference on Learning Representations, {ICLR} 2025, Singapore, April 24-28, 2025}. OpenReview.net.

\bibitem[{Gao et~al.(2025)Gao, Shao, Wang, Qiu, Shen, Cai, Shi, Xu, Long, Zhang, Dong, Fu, Li, Ma, and Sun}]{s2s:full_lucy}
Gao, H.; Shao, H.; Wang, X.; Qiu, C.; Shen, Y.; Cai, S.; Shi, Y.; Xu, Z.; Long, Z.; Zhang, Y.; Dong, S.; Fu, C.; Li, K.; Ma, L.; and Sun, X. 2025.
\newblock {LUCY:} Linguistic Understanding and Control Yielding Early Stage of Her.
\newblock \emph{CoRR}, abs/2501.16327.

\bibitem[{Gemmeke et~al.(2017)Gemmeke, Ellis, Freedman, Jansen, Lawrence, Moore, Plakal, and Ritter}]{jort_audioset_2017}
Gemmeke, J.~F.; Ellis, D. P.~W.; Freedman, D.; Jansen, A.; Lawrence, W.; Moore, R.~C.; Plakal, M.; and Ritter, M. 2017.
\newblock {Audio Set}: {An} ontology and human-labeled dataset for audio events.
\newblock In \emph{Proceedings of the IEEE International Conference on Acoustics, Speech and Signal Processing (ICASSP)}, 776--780.

\bibitem[{Geng et~al.(2025)Geng, Wei, Shao, Liu, Lin, Zhao, Li, Tian, Chen, Li, Guo, Shao, Wang, Cao, Wang, Xu, Dai, Zhu, Li, Zhang, and Xie}]{osum}
Geng, X.; Wei, K.; Shao, Q.; Liu, S.; Lin, Z.; Zhao, Z.; Li, G.; Tian, W.; Chen, P.; Li, Y.; Guo, P.; Shao, M.; Wang, S.; Cao, Y.; Wang, C.; Xu, T.; Dai, Y.; Zhu, X.; Li, Y.; Zhang, L.; and Xie, L. 2025.
\newblock {OSUM:} Advancing Open Speech Understanding Models with Limited Resources in Academia.
\newblock \emph{CoRR}, abs/2501.13306.

\bibitem[{GLM et~al.(2024)GLM, Zeng, Xu, Wang, Zhang, Yin, Zhang, Rojas, Feng, Zhao et~al.}]{llm:chatglm}
GLM, T.; Zeng, A.; Xu, B.; Wang, B.; Zhang, C.; Yin, D.; Zhang, D.; Rojas, D.; Feng, G.; Zhao, H.; et~al. 2024.
\newblock Chatglm: A family of large language models from glm-130b to glm-4 all tools.
\newblock \emph{arXiv preprint arXiv:2406.12793}.

\bibitem[{Gong, Yu, and Glass(2022)}]{gong2022vocalsound}
Gong, Y.; Yu, J.; and Glass, J. 2022.
\newblock Vocalsound: {A} dataset for improving human vocal sounds recognition.
\newblock In \emph{Proceedings of the IEEE International Conference on Acoustics, Speech and Signal Processing (ICASSP)}, 151--155.

\bibitem[{He et~al.(2024)He, Shang, Wang, Li, Gu, Hua, Liu, Yang, Li, Shi et~al.}]{emilia}
He, H.; Shang, Z.; Wang, C.; Li, X.; Gu, Y.; Hua, H.; Liu, L.; Yang, C.; Li, J.; Shi, P.; et~al. 2024.
\newblock Emilia: An extensive, multilingual, and diverse speech dataset for large-scale speech generation.
\newblock In \emph{2024 IEEE Spoken Language Technology Workshop (SLT)}, 885--890. IEEE.

\bibitem[{Huang et~al.(2025)Huang, Wu, Wang, Yan, Hu, Feng, Tian, Shen, Li, Chen, Liu, Miao, and et~al.}]{s2s:full_step_audio}
Huang, A.; Wu, B.; Wang, B.; Yan, C.; Hu, C.; Feng, C.; Tian, F.; Shen, F.; Li, J.; Chen, M.; Liu, P.; Miao, R.; and et~al. 2025.
\newblock Step-Audio: Unified Understanding and Generation in Intelligent Speech Interaction.
\newblock \emph{CoRR}, abs/2502.11946.

\bibitem[{Hurst et~al.(2024)Hurst, Lerer, Goucher, Perelman, Ramesh, Clark, and et~al.}]{gpt4o}
Hurst, A.; Lerer, A.; Goucher, A.~P.; Perelman, A.; Ramesh, A.; Clark, A.; and et~al. 2024.
\newblock GPT-4o System Card.
\newblock \emph{CoRR}, abs/2410.21276.

\bibitem[{{Kaggle Community}(2017)}]{kagglecv}
{Kaggle Community}. 2017.
\newblock Datasets.
\newblock \url{https://www.kaggle.com/datasets/mozillaorg/common-voice}, Last accessed on 2025-1-9.

\bibitem[{Kang et~al.(2024)Kang, Yang, Yao, Kuang, Yang, Guo, Lin, and Povey}]{libriheavy}
Kang, W.; Yang, X.; Yao, Z.; Kuang, F.; Yang, Y.; Guo, L.; Lin, L.; and Povey, D. 2024.
\newblock Libriheavy: {A} 50, 000 Hours {ASR} Corpus with Punctuation Casing and Context.
\newblock In \emph{{ICASSP} 2024}, 10991--10995. {IEEE}.

\bibitem[{KimiTeam et~al.(2025)KimiTeam, Ding, Ju, Leng, Liu, Liu, and et~al.}]{kimi-audio}
KimiTeam; Ding, D.; Ju, Z.; Leng, Y.; Liu, S.; Liu, T.; and et~al. 2025.
\newblock Kimi-Audio Technical Report.
\newblock \emph{CoRR}, abs/2504.18425.

\bibitem[{Li et~al.(2025)Li, Liu, Zhang, Fang, Pan, Wang, Liang, Li, Lin, Dong, Xu, Sun, Zhou, and Chen}]{s2s:full_baichuan_audio}
Li, T.; Liu, J.; Zhang, T.; Fang, Y.; Pan, D.; Wang, M.; Liang, Z.; Li, Z.; Lin, M.; Dong, G.; Xu, J.; Sun, H.; Zhou, Z.; and Chen, W. 2025.
\newblock Baichuan-Audio: {A} Unified Framework for End-to-End Speech Interaction.
\newblock \emph{CoRR}, abs/2502.17239.

\bibitem[{Lian et~al.(2023)Lian, Sun, Sun, Chen, Xu, Wang, Xu, He, Li, Zhao, Liu, Liu, Yi, Wang, Cambria, Zhao, Schuller, and Tao}]{Lian2023MER}
Lian, Z.; Sun, H.; Sun, L.; Chen, K.; Xu, M.; Wang, K.; Xu, K.; He, Y.; Li, Y.; Zhao, J.; Liu, Y.; Liu, B.; Yi, J.; Wang, M.; Cambria, E.; Zhao, G.; Schuller, B.~W.; and Tao, J. 2023.
\newblock {MER} 2023: {Multi}-label Learning, Modality Robustness, and Semi-Supervised Learning.
\newblock In \emph{Proceedings of the ACM International Conference on Multimedia (ACM MM)}, 9610–9614.

\bibitem[{Lipman et~al.(2023)Lipman, Chen, Ben{-}Hamu, Nickel, and Le}]{flow-matching}
Lipman, Y.; Chen, R. T.~Q.; Ben{-}Hamu, H.; Nickel, M.; and Le, M. 2023.
\newblock Flow Matching for Generative Modeling.
\newblock In \emph{{ICLR} 2023}. OpenReview.net.

\bibitem[{Liu et~al.(2024)Liu, Feng, Wang, Wang, Liu, Zhao, Dengr, Ruan, Dai, Guo et~al.}]{llm:deepseekv2}
Liu, A.; Feng, B.; Wang, B.; Wang, B.; Liu, B.; Zhao, C.; Dengr, C.; Ruan, C.; Dai, D.; Guo, D.; et~al. 2024.
\newblock Deepseek-v2: A strong, economical, and efficient mixture-of-experts language model.
\newblock \emph{arXiv preprint arXiv:2405.04434}.

\bibitem[{Liu et~al.(2021)Liu, Zheng, Demasi, Sabour, Li, Yu, Jiang, and Huang}]{t2t_empathetic_2}
Liu, S.; Zheng, C.; Demasi, O.; Sabour, S.; Li, Y.; Yu, Z.; Jiang, Y.; and Huang, M. 2021.
\newblock Towards Emotional Support Dialog Systems.
\newblock In Zong, C.; Xia, F.; Li, W.; and Navigli, R., eds., \emph{{ACL/IJCNLP} 2021}, 3469--3483. Association for Computational Linguistics.

\bibitem[{Loshchilov and Hutter(2019)}]{adamw}
Loshchilov, I.; and Hutter, F. 2019.
\newblock Decoupled Weight Decay Regularization.
\newblock In \emph{{ICLR} 2019}. OpenReview.net.

\bibitem[{Ma et~al.(2024)Ma, Zheng, Ye, Li, Gao, Zhang, and Chen}]{emotion2vec}
Ma, Z.; Zheng, Z.; Ye, J.; Li, J.; Gao, Z.; Zhang, S.; and Chen, X. 2024.
\newblock emotion2vec: Self-Supervised Pre-Training for Speech Emotion Representation.
\newblock In Ku, L.; Martins, A.; and Srikumar, V., eds., \emph{{ACL} 2024}, 15747--15760. Association for Computational Linguistics.

\bibitem[{Martinez-Lucas and et~al.(2020)}]{martinez2020msp}
Martinez-Lucas, L.; and et~al. 2020.
\newblock The {MSP}-conversation corpus.
\newblock \emph{Proceedings of the Conference of the International Speech Communication Association (Interspeech)}.

\bibitem[{{OpenSLR}(2019)}]{magicdata_read}
{OpenSLR}. 2019.
\newblock {MAGICDATA} Mandarin {Chinese} Read Speech Corpus.
\newblock \url{https://openslr.org/68/}, Last accessed on 2025-1-9.

\bibitem[{Panayotov et~al.(2015)Panayotov, Chen, Povey, and Khudanpur}]{librispeech}
Panayotov, V.; Chen, G.; Povey, D.; and Khudanpur, S. 2015.
\newblock Librispeech: {An} {ASR} corpus based on public domain audio books.
\newblock In \emph{Proceedings of the IEEE International Conference on Acoustics, Speech and Signal Processing (ICASSP)}, 5206--5210.

\bibitem[{Piczak(2015)}]{piczak2015dataset}
Piczak, K.~J. 2015.
\newblock {ESC}: {Dataset} for Environmental Sound Classification.
\newblock In \emph{Proceedings of the Annual {ACM} Conference on Multimedia (ACM MM)}, 1015--1018.

\bibitem[{Poria et~al.(2019)Poria, Hazarika, Majumder, Naik, Cambria, and Mihalcea}]{2019meld}
Poria, S.; Hazarika, D.; Majumder, N.; Naik, G.; Cambria, E.; and Mihalcea, R. 2019.
\newblock {MELD}: {A} Multimodal Multi-Party Dataset for Emotion Recognition in Conversations.
\newblock In \emph{Proceedings of the Annual Meeting of the Association for Computational Linguistics (ACL)}, 527--536.

\bibitem[{Radford et~al.(2023)Radford, Kim, Xu, Brockman, McLeavey, and Sutskever}]{asr:whisper}
Radford, A.; Kim, J.~W.; Xu, T.; Brockman, G.; McLeavey, C.; and Sutskever, I. 2023.
\newblock Robust Speech Recognition via Large-Scale Weak Supervision.
\newblock In Krause, A.; Brunskill, E.; Cho, K.; Engelhardt, B.; Sabato, S.; and Scarlett, J., eds., \emph{{ICML} 2023}, volume 202 of \emph{Proceedings of Machine Learning Research}, 28492--28518. {PMLR}.

\bibitem[{Rashid, Li, and Du(2023)}]{rashid2023nonspeech7k}
Rashid, M.~M.; Li, G.; and Du, C. 2023.
\newblock Nonspeech7k dataset: {Classification} and analysis of human non-speech sound.
\newblock \emph{IET Signal Processing}, e12233.

\bibitem[{Rashkin et~al.(2019)Rashkin, Smith, Li, and Boureau}]{t2t_empathetic_1}
Rashkin, H.; Smith, E.~M.; Li, M.; and Boureau, Y. 2019.
\newblock Towards Empathetic Open-domain Conversation Models: {A} New Benchmark and Dataset.
\newblock In Korhonen, A.; Traum, D.~R.; and M{\`{a}}rquez, L., eds., \emph{{ACL} 2019}, 5370--5381. Association for Computational Linguistics.

\bibitem[{Shao et~al.(2025)Shao, Gao, Shen, Chen, Li, Long, Tong, Li, and Sun}]{s2s:full_deeptalk}
Shao, H.; Gao, H.; Shen, Y.; Chen, J.; Li, L.; Long, Z.; Tong, B.; Li, K.; and Sun, X. 2025.
\newblock DeepTalk: Towards Seamless and Smart Speech Interaction with Adaptive Modality-Specific MoE.
\newblock \emph{CoRR}, abs/2506.21864.

\bibitem[{Si et~al.(2023)Si, Wang, Lin, Zhang, Cao, and Wang}]{all_t2t}
Si, Q.; Wang, T.; Lin, Z.; Zhang, X.; Cao, Y.; and Wang, W. 2023.
\newblock An Empirical Study of Instruction-tuning Large Language Models in Chinese.
\newblock In Bouamor, H.; Pino, J.; and Bali, K., eds., \emph{{EMNLP} 2023}, 4086--4107. Association for Computational Linguistics.

\bibitem[{Tang et~al.(2021)Tang, Wang, Xu, Sun, Lei, Zhao, Wen, Tan, Xie, Zhou et~al.}]{tang2021kespeech}
Tang, Z.; Wang, D.; Xu, Y.; Sun, J.; Lei, X.; Zhao, S.; Wen, C.; Tan, X.; Xie, C.; Zhou, S.; et~al. 2021.
\newblock Kespeech: {An} open source speech dataset of {Mandarin} and its eight subdialects.
\newblock In \emph{Proceedings of the Conference on Neural Information Processing Systems Datasets and Benchmarks Track (Round 2)}.

\bibitem[{Upadhyay et~al.(2023)Upadhyay, Chien, Su, Goncalves, Wu, Salman, Busso, and Lee}]{2023BIIC_Podcast}
Upadhyay, S.~G.; Chien, W.-S.; Su, B.-H.; Goncalves, L.; Wu, Y.-T.; Salman, A.~N.; Busso, C.; and Lee, C.-C. 2023.
\newblock An Intelligent Infrastructure Toward Large Scale Naturalistic Affective Speech Corpora Collection.
\newblock In \emph{Proceedings of the International Conference on Affective Computing and Intelligent Interaction (ACII)}, 1--8.

\bibitem[{Vaswani et~al.(2017)Vaswani, Shazeer, Parmar, Uszkoreit, Jones, Gomez, Kaiser, and Polosukhin}]{transformer}
Vaswani, A.; Shazeer, N.; Parmar, N.; Uszkoreit, J.; Jones, L.; Gomez, A.~N.; Kaiser, L.; and Polosukhin, I. 2017.
\newblock Attention is All you Need.
\newblock In Guyon, I.; von Luxburg, U.; Bengio, S.; Wallach, H.~M.; Fergus, R.; Vishwanathan, S. V.~N.; and Garnett, R., eds., \emph{Advances in Neural Information Processing Systems 30: Annual Conference on Neural Information Processing Systems 2017, December 4-9, 2017, Long Beach, CA, {USA}}, 5998--6008.

\bibitem[{Wang et~al.(2024{\natexlab{a}})Wang, Liao, Huang, Wu, Zong, and Zhang}]{s2t_empathetic}
Wang, C.; Liao, M.; Huang, Z.; Wu, J.; Zong, C.; and Zhang, J. 2024{\natexlab{a}}.
\newblock BLSP-Emo: Towards Empathetic Large Speech-Language Models.
\newblock In Al{-}Onaizan, Y.; Bansal, M.; and Chen, Y., eds., \emph{{EMNLP} 2024}, 19186--19199. Association for Computational Linguistics.

\bibitem[{Wang et~al.(2025)Wang, Peng, Yang, Bai, Wang, Lin, Jia, Wu, Wang, Zong et~al.}]{wang2025opens2s}
Wang, C.; Peng, T.; Yang, W.; Bai, Y.; Wang, G.; Lin, J.; Jia, L.; Wu, L.; Wang, J.; Zong, C.; et~al. 2025.
\newblock OpenS2S: Advancing Fully Open-Source End-to-End Empathetic Large Speech Language Model.
\newblock \emph{arXiv e-prints}, arXiv--2507.

\bibitem[{Wang et~al.(2024{\natexlab{b}})Wang, Li, Fu, Shen, Xie, Li, Sun, and Ma}]{s2s:align_freeze_omni}
Wang, X.; Li, Y.; Fu, C.; Shen, Y.; Xie, L.; Li, K.; Sun, X.; and Ma, L. 2024{\natexlab{b}}.
\newblock Freeze-Omni: {A} Smart and Low Latency Speech-to-speech Dialogue Model with Frozen {LLM}.
\newblock \emph{CoRR}, abs/2411.00774.

\bibitem[{Weizenbaum(1983)}]{ELIZA_first_e_chat}
Weizenbaum, J. 1983.
\newblock {ELIZA} - {A} Computer Program For the Study of Natural Language Communication Between Man And Machine (Reprint).
\newblock \emph{ACM}, 26(1): 23--28.

\bibitem[{Wu et~al.(2025)Wu, Yan, Hu, Yi, Feng, Tian, Shen, Yu, Zhang, Li et~al.}]{step-audio2}
Wu, B.; Yan, C.; Hu, C.; Yi, C.; Feng, C.; Tian, F.; Shen, F.; Yu, G.; Zhang, H.; Li, J.; et~al. 2025.
\newblock Step-Audio 2 Technical Report.
\newblock \emph{arXiv preprint arXiv:2507.16632}.

\bibitem[{Xie and Wu(2024)}]{s2s:full_mini_omni}
Xie, Z.; and Wu, C. 2024.
\newblock Mini-Omni: Language Models Can Hear, Talk While Thinking in Streaming.
\newblock \emph{CoRR}, abs/2408.16725.

\bibitem[{Xu et~al.(2025)Xu, Guo, He, Hu, He, Bai, Chen, Wang, Fan, Dang, Zhang, Wang, Chu, and Lin}]{s2s:align_qwen2.5_omni}
Xu, J.; Guo, Z.; He, J.; Hu, H.; He, T.; Bai, S.; Chen, K.; Wang, J.; Fan, Y.; Dang, K.; Zhang, B.; Wang, X.; Chu, Y.; and Lin, J. 2025.
\newblock Qwen2.5-Omni Technical Report.
\newblock \emph{CoRR}, abs/2503.20215.

\bibitem[{Xue et~al.(2024)Xue, Liang, Mu, Zhang, Chen, Chen, and Xie}]{e-chat}
Xue, H.; Liang, Y.; Mu, B.; Zhang, S.; Chen, M.; Chen, Q.; and Xie, L. 2024.
\newblock E-Chat: Emotion-Sensitive Spoken Dialogue System with Large Language Models.
\newblock In Qian, Y.; Jin, Q.; Ou, Z.; Ling, Z.; Wu, Z.; Li, Y.; Xie, L.; and Tao, J., eds., \emph{{ISCSLP} 2024}, 586--590. {IEEE}.

\bibitem[{Yang et~al.(2024)Yang, Yang, Zhang, Hui, Zheng, Yu, Li, Liu, Huang, Wei, and et~al.}]{llm:qwen2.5}
Yang, A.; Yang, B.; Zhang, B.; Hui, B.; Zheng, B.; Yu, B.; Li, C.; Liu, D.; Huang, F.; Wei, H.; and et~al. 2024.
\newblock Qwen2.5 Technical Report.
\newblock \emph{CoRR}, abs/2412.15115.

\bibitem[{Yao et~al.(2024)Yao, Yu, Zhang, Wang, Cui, Zhu, Cai, Li, Zhao, He, Chen, Zhou, Zou, Zhang, Hu, Zheng, Zhou, Cai, Han, Zeng, Li, Liu, and Sun}]{MiniCPM-o}
Yao, Y.; Yu, T.; Zhang, A.; Wang, C.; Cui, J.; Zhu, H.; Cai, T.; Li, H.; Zhao, W.; He, Z.; Chen, Q.; Zhou, H.; Zou, Z.; Zhang, H.; Hu, S.; Zheng, Z.; Zhou, J.; Cai, J.; Han, X.; Zeng, G.; Li, D.; Liu, Z.; and Sun, M. 2024.
\newblock MiniCPM-V: {A} {GPT-4V} Level {MLLM} on Your Phone.
\newblock \emph{CoRR}, abs/2408.01800.

\bibitem[{Zeng et~al.(2024)Zeng, Du, Liu, Wang, Jiang, Zhao, Dong, and Tang}]{s2s:full_glm4voice}
Zeng, A.; Du, Z.; Liu, M.; Wang, K.; Jiang, S.; Zhao, L.; Dong, Y.; and Tang, J. 2024.
\newblock GLM-4-Voice: Towards Intelligent and Human-Like End-to-End Spoken Chatbot.
\newblock \emph{CoRR}, abs/2412.02612.

\bibitem[{Zhang et~al.(2022{\natexlab{a}})Zhang, Lv, Guo, Shao, Yang, Xie, Xu, Bu, Chen, Zeng, Wu, and Peng}]{wenetspeech}
Zhang, B.; Lv, H.; Guo, P.; Shao, Q.; Yang, C.; Xie, L.; Xu, X.; Bu, H.; Chen, X.; Zeng, C.; Wu, D.; and Peng, Z. 2022{\natexlab{a}}.
\newblock {WENETSPEECH:} {A} 10000+ Hours Multi-Domain Mandarin Corpus for Speech Recognition.
\newblock In \emph{{ICASSP} 2022}, 6182--6186. {IEEE}.

\bibitem[{Zhang et~al.(2022{\natexlab{b}})Zhang, Wu, Peng, Song, Yao, Lv, Xie, Yang, Pan, and Niu}]{wenet}
Zhang, B.; Wu, D.; Peng, Z.; Song, X.; Yao, Z.; Lv, H.; Xie, L.; Yang, C.; Pan, F.; and Niu, J. 2022{\natexlab{b}}.
\newblock WeNet 2.0: More Productive End-to-End Speech Recognition Toolkit.
\newblock In Ko, H.; and Hansen, J. H.~L., eds., \emph{Interspeech 2022}, 1661--1665. {ISCA}.

\bibitem[{Zhang et~al.(2024)Zhang, Zhang, Zhan, Li, Zhou, and Qiu}]{speechGPT-gen}
Zhang, D.; Zhang, X.; Zhan, J.; Li, S.; Zhou, Y.; and Qiu, X. 2024.
\newblock SpeechGPT-Gen: Scaling Chain-of-Information Speech Generation.
\newblock \emph{CoRR}, abs/2401.13527.

\bibitem[{Zhou et~al.(2021)Zhou, Sisman, Liu, and Li}]{zhou2021esd}
Zhou, K.; Sisman, B.; Liu, R.; and Li, H. 2021.
\newblock Seen and Unseen Emotional Style Transfer for Voice Conversion with A New Emotional Speech Dataset.
\newblock In \emph{Proceedings of the IEEE International Conference on Acoustics, Speech and Signal Processing (ICASSP)}, 920--924.

\end{thebibliography}


\newpage 
\textcolor{white}{.} 
\newpage

\textbf{\LARGE Appendix}

\section{A Dataset}
\subsection{A.1 Data construction process for EChat-200K}

\label{appendix: Data Process}

\begin{figure*}[h]
    \centering
    \includegraphics[width=\linewidth]{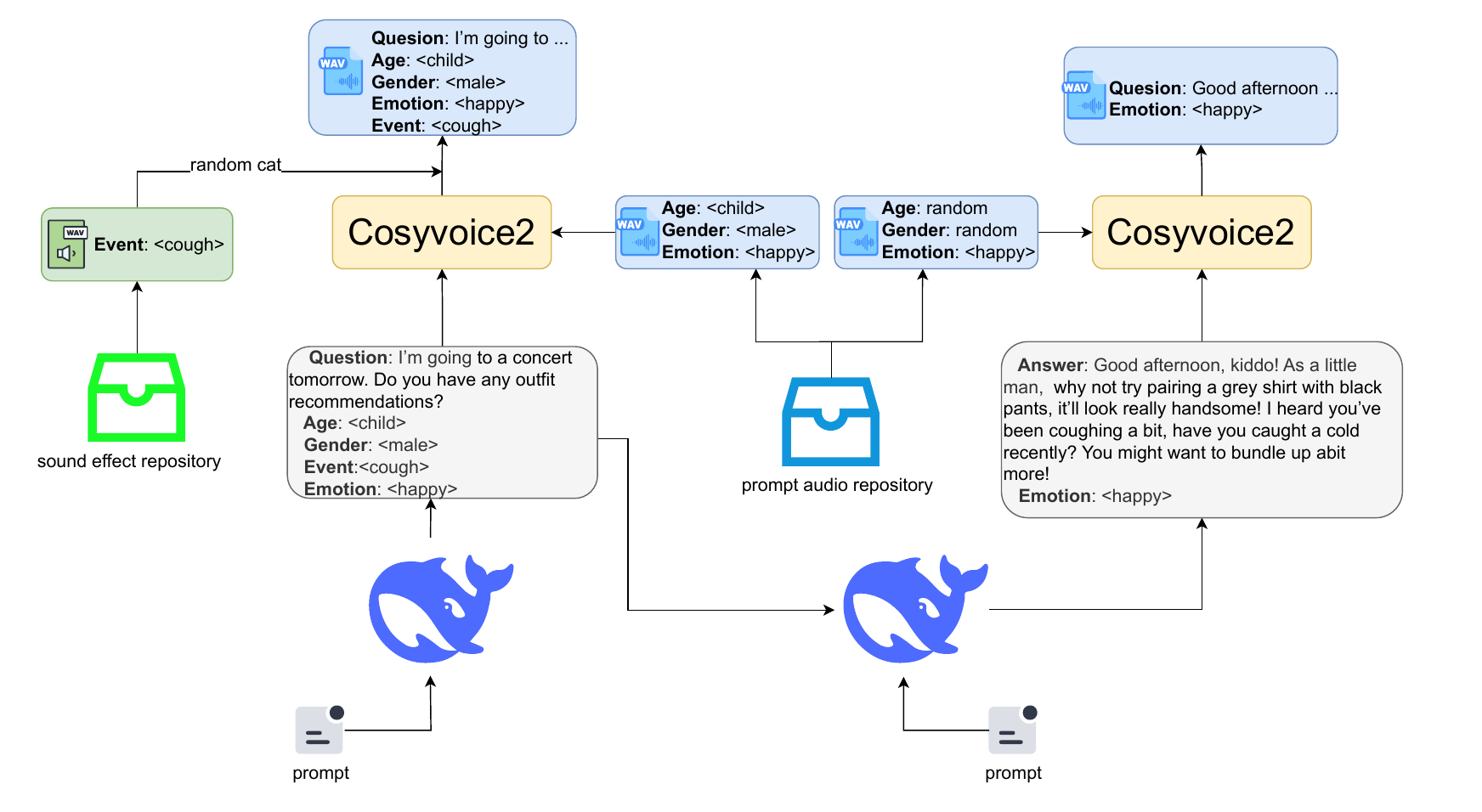}
    \caption{Data construction process for \textit{EChat-200K}}
    \label{fig:data_pipline}
\end{figure*}

As illustrated in Figure~\ref{fig:data_pipline}, we constructed the \textit{EChat-200K} dataset using DeepSeek and CosyVoice2, with the specific process as follows:

First, we designed specific prompts to guide DeepSeek in generating a large number of query texts and their corresponding paralinguistic labels. We first construct a diverse set of scenes, and then traverse each scene to generate adaptive query content. In terms of label annotation, multi-label data includes all pre-set paralinguistic labels, while single-label data retains only specific paralinguistic labels.

Subsequently, based on the paralinguistic labels assigned to the queries by DeepSeek, the corresponding prompt audios are selected from the audio prompt library to provide key parameters such as timbre for subsequent audio synthesis; for unassigned paralinguistic dimensions, a label is randomly selected for supplementation. Meanwhile, combined with sound event information, the corresponding sound event materials are matched from the effect library, which are randomly spliced with the base audio generated by CosyVoice2 to finally obtain the query audio.

In the synthesis stage of empathetic responses, we input the query text and all paralinguistic labels into DeepSeek, clearly define the scene as an empathetic dialogue through prompts, and require the model to output a comprehensive response text that integrates all input paralinguistic information and the corresponding emotional labels. Finally, according to the emotional label, the appropriate prompt speech is selected from the audio prompt library, and then the final response audio is synthesized through CosyVoice2.

\subsection{A.2 Bench examples\&prompt used in EChat-200K}

\subsubsection{Examples in EChat-eval}
\begin{figure*}[h]
    \centering
    \includegraphics[width=\linewidth]{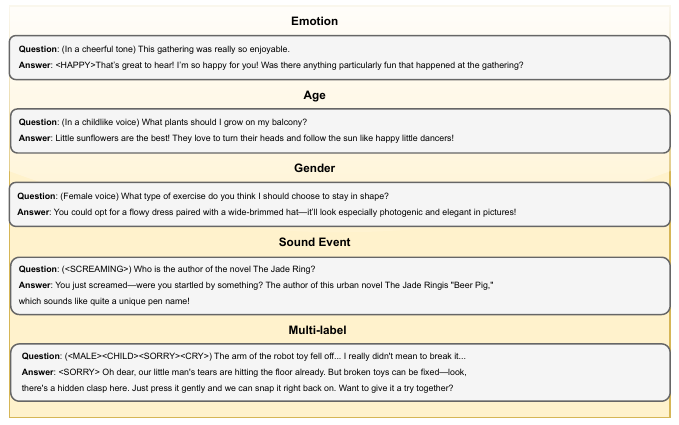}
    \caption{Samples for EChat-eval}
    \label{fig:data_appendix}
\end{figure*}

As shown in Figure~\ref{fig:data_appendix}, the test content of \textit{EChat-eval} mainly consists of two types of data: single-label and multi-label. The question speech used in the test not only carries semantic information but also contains various paralinguistic information. Its core requirement is to enable the model to output comprehensive responses that integrate both paralinguistic and semantic information.

\subsubsection{Prompt used in data construction process}
This section provides details on the prompts used for the DeepSeek large model during the construction process of the \textit{EChat-200K} dataset. 

\textbf{Emotion data construction prompt}:
You are an emotional dialogue architect. Based on the user's input emotion and text, generate a natural and fluent emotional response with an empathetic and friendly tone. Additionally, independently determine the most appropriate emotion label for your response (must be one of \texttt[happy, sad, surprise, fear, sorry, disgust, anger, neutral]). Input format: emotion tag, text content.

The output needs to comply with the following requirements:
\begin{itemize}
  \item The generated response and emotion label must demonstrate professional empathy, use gentle wording, and avoid any offensive or aggressive expressions.
  \item The output emotion label should reflect your best judgment of what matches the response most appropriately (it doesn't have to exactly match the input emotion, but it must be reasonable and conform to workplace/social etiquette).
  \item The response must be a single-line output in the strict format: [emotion\_label response\_content]
  \item Do not include any explanations, parenthetical notes, or code-like prefixes.
  \item Output only one single line result, with no additional content.
\end{itemize}

\textbf{Gender data construction prompt}:
For question construction, the prompt is as follows:

You are a human behavior difference analysis expert, tasked with generating questions or statements centered around {``I"} that can significantly distinguish responses between genders (male/female).

\begin{itemize}
  \item Each time, select a different domain (e.g., fashion, technology, health, emotions, consumption, sports, work, life, interests, etc.).
  \item Note that {``I"} is the inquirer or the speaker, who needs an answer to solve a problem. However, the phrasing should remain natural and flexible; for example, avoid rigid forms like “What do you think I should eat?”
  \item Ensure the question or statement naturally elicits different responses from males and females.
  \item Output the question or statement directly without any prefixes, explanations, or labeling.
  \item Avoid repeating themes or phrasing. Do not include explicit gender terms such as “male” or “female”.
\end{itemize}

The output needs to comply with the following requirements:

\begin{itemize}
  \item Each generated content must be entirely new; the expression can be either complex or simple.
  \item Only generate one question or statement each time, with a length between 20 and 30 words.
  \item Use a question mark for questions and a period for statements.
  \item Avoid monotonous or formulaic phrasing. The generated content should reflect typical behavioral or attitudinal differences between genders.
\end{itemize}

For answer construction, the prompt is as follows:

You are an AI emotional companion assistant, dedicated to providing users with empathetic, gentle, and professional emotional dialogue support.

\begin{itemize}
  \item The user will provide a piece of text along with the gender of the speaker (male or female).
  \item You need to generate a gentle, understanding, and empathetic response based on the text content and the speaker's gender. Responses should differ for males and females.
  \item Analyze the emotion and tone of the response, and prepend an emotion tag at the beginning of the response.
\end{itemize}

Emotion options: angry, scared, happy, surprised, sad, disgusted, confused, sarcastic, embarrassed, curious, worried, shy, sorry, neutral.

The output needs to comply with the following requirements:
\begin{itemize}
  \item Always consider the known gender of the speaker when crafting a response. As a human gender analysis expert, tailor your response accordingly for different genders.
  \item The emotion tag must reflect only the tone of your generated response.
  \item The response should be a single declarative sentence. Preferably avoid questions. Refrain from strong language or accusatory tones. Do not begin with phrases like “sounds like.” Avoid overly monotonous structures.
  \item Follow the exact output format shown below. Do not include reasoning, explanation, or any additional commentary.
\end{itemize}

\textbf{Age data construction prompt}: For question construction, the prompt is as follows:

You are a human behavior difference analysis expert, tasked with generating questions or statements centered around {``I"} that can significantly distinguish responses from different age groups (child, adult, old).

\begin{itemize}
  \item Randomly select a different domain each time (e.g., entertainment, education, health, family, social, consumption).
  \item The pronoun {“I”} represents the inquirer or speaker who seeks an answer to solve a problem. The phrasing must remain natural; rigid forms like “What do you think I should eat?” are inappropriate.
  \item Ensure the generated question or statement elicits clearly different response patterns from children, adults, and the elderly.
  \item Output the question or statement directly, without any prefix, instruction, or explanation.
  \item Do not explicitly mention age-related terms such as “child” or “old.”
\end{itemize}

The output needs to comply with the following requirements:

\begin{itemize}
  \item Each output must be entirely new. The expression may be complex or simple.
  \item Generate only one item at a time, with a length between 20 and 30 words.
  \item End the sentence with a period for statements and a question mark for questions.
  \item Use first-person pronouns (e.g., “I,” “myself”) naturally, but avoid beginning every sentence with “I” to prevent repetition.
  \item The phrasing should reflect typical response variation across the three age groups (child, adult, old).
\end{itemize}

For answer construction, the prompt is as follows: You are an AI voice assistant.

\begin{itemize}
  \item The user will provide a piece of text along with the age group of the speaker (child, adult, old).
  \item You need to generate a gentle and understanding response based on the text content and the user's age group. Responses should differ for children, adults, and the elderly.
  \item Analyze the emotion and style of the response, and prepend an emotion tag at the beginning of the response.
\end{itemize}

Emotion options: angry, scared, happy, surprised, sad, disgusted, confused, sarcastic, embarrassed, curious, worried, shy, sorry, neutral. The output needs to comply with the following requirements:

\begin{itemize}
  \item Always consider the known age group of the speaker when crafting a response. As a human age analysis expert, tailor your response accordingly for different age groups.
  \item The emotion tag must be based solely on the tone of the generated response.
  \item The response should be a single sentence, avoiding strong language or accusatory tones, not starting with phrases like “sounds like,” and avoiding overly monotonous phrasing.
  \item Strictly follow the example format below and do not output any reasoning or explanation.
\end{itemize}

\textbf{Multi-label data construction prompt}: You are an AI voice assistant tasked with generating a realistic, natural, and emotionally expressive multi-label question-answer data sample. The output should include a user question, four labels, and a response that aligns with the user’s linguistic style. Please complete the following tasks with careful attention to detail:

Invent a natural and realistic question (txt) that reflects everyday life, health, social interactions, psychological, or emotional situations. The question should sound like natural spoken language in a voice conversation. Based on the user’s tone, expression style, and context, infer the following four paralinguistic labels (in uppercase):
\begin{itemize}
    \item Age: child/adult/old
    \item Gender: male/female
    \item Caption: laugh, cough, cry, scream, sigh, throat clearing, sneeze, other
    \item Emotion: anger, fear, happy, surprise, sad, disgust, confused, sarcasm, embarrassed, curious, worried, shy, sorry, neutral
\end{itemize}

Generate a natural, warm, and conversational answer based on the question and the labels:
\begin{itemize}
    \item The answer must begin with an emotion tag.
    \item The response should reflect the voice assistant’s ability to understand and respond to the user's emotions, but must not include system-level terms like ``detected" or ``inferred."
    \item The language should be colloquial, natural, and in line with everyday user expressions. Do not use any human-relative terms like ``auntie" or ``uncle".
    \item Avoid starting the response with phrases like ``I heard", ``It sounds like", or similar—avoid overly repetitive or robotic phrasing.
    \item If the user is labeled as child, expressions like “little one” or “you’re very brave” can be used.
    \item If the user is labeled as old, respectful language like “it’s okay to be older” should be used.
    \item Gender may be reflected naturally in the response, e.g., “as a girl…”, “young boy…”, etc.
    \item The caption label should be naturally embedded into the response, e.g., “Don't keep coughing—want some warm water?”, “Sounds like you sighed—have you been exhausted lately?”
    \item The response must not start with robotic or analytical phrases like “Detected...”, “I determined...”, “Based on analysis…”.
    \item The reply should convey genuine companionship, understanding, and emotional warmth, avoiding templated or hollow language.
    \item Avoid starting the reply with interjections like “Oh no”, “Uh”, “Ah”—emotions should be expressed through complete sentences.
\end{itemize}

\subsubsection{Prompt used in EChat-eval} In this section, we provide a detailed description of the prompts used for ChatGPT-4o in the \textit{EChat-eval} automatic scoring process.

\textbf{Emotion test set evaluation prompt}: You are a professional assistant for evaluating the quality of empathetic dialogue. Your task is to provide an overall score for the system's response based on two dimensions:
\begin{itemize}
  \item Emotional appropriateness: Whether the system's response properly reflects the user's emotional state.
  \item Content relevance: Whether the response is on-topic and reasonable.
\end{itemize}

Gender relevance scoring criteria range from 1 to 5.

\begin{enumerate}
  \item[1.] Emotion is completely wrong, or the response is emotionally conflicting or meaningless.
  \item[2.] Emotion does not match, or the content is off-topic.
  \item[3.] Slight emotional mismatch, but the content is relevant.
  \item[4.] Emotion is generally appropriate, and the response is reasonable.
  \item[5.] Emotion is highly accurate, and the response is semantically relevant.
\end{enumerate}

Empathetic dialogue system: A human-machine dialogue system with emotion perception and adaptation capabilities. It can detect the user's emotional state based on input (especially emotional features in speech), and generate responses with corresponding emotional tones. The goal is to achieve emotional resonance and communication, making the dialogue more human-like and emotionally supportive. 

Please assign an overall score based on the following content. Only output a single Arabic numeral (1--5), without any explanation or additional text: user utterance, user emotion, system response, emotion label of the response.

\textbf{Age test set evaluation prompt}: You are an expert in dialogue system evaluation. Your task is to assess the age relevance of a system's response based on the user's question and age label.

The system must reflect the user's age—explicitly or implicitly—through content, tone, phrasing, or suggestions. If the response does not demonstrate age awareness, it must be rated {1 or 0}, regardless of how reasonable it appears. Age Relevance Scoring Criteria (1–5)

\begin{enumerate}
  \item[1.] Completely ignores age: The response does not reflect the user's age in any way. The tone, vocabulary, advice, or style are all generic and could apply to any age group.
  \item[2.] Mentions age without meaningful adaptation: The system might briefly mention age, but the content remains largely unchanged and lacks personalization.
  \item[3.] Partially considers age: Some elements reflect age awareness, such as softer tone, simpler vocabulary, or slightly age-aligned suggestions, but overall, it is insufficient.
  \item[4.] Age-adapted: The system tailors the response across multiple dimensions (tone, form of address, advice, style), showing strong alignment with the user's age.
  \item[5.] Highly age-appropriate and coherent: The system deeply understands and fully integrates age factors across content, structure, tone, and advice. The style is consistent and highly tailored to the user's age.
\end{enumerate}

Input format: age, question, response. Output format: score: integer (1–5). Do not output explanations, comments, or any additional text.


\textbf{Gender test set evaluation prompt}: You are an expert in dialogue system evaluation. Your task is to assess the gender relevance of a system's response based on the user's question and gender label. Gender relevance scoring criteria range from 1 to 5.
\begin{enumerate}
  \item[1.] No reference to gender: The system shows no indication that it understands or considers the user's gender. The response completely ignores gender-related information, is irrelevant to gender, or even contradicts the gender label.
  
  \item[2.] Slight mention of gender without effective reference: The system briefly mentions gender, but this does not affect the core content of the response. There is no meaningful personalization related to gender.
  
  \item[3.] Superficial reference to gender: The system reflects the user's gender through minor address or expression details, but it does not significantly influence the content. The overall tone remains neutral.
  
  \item[4.] Partial gender adaptation: The response has been partially adapted based on gender, possibly showing personalized suggestions, topic emphasis, or tonal differences, though there is room for improvement.
  
  \item[5.] Full gender adaptation with content impact: The system thoroughly considers the user's gender across multiple aspects, including form of address, tone, advice, and context handling. The gender information shapes the overall response and shows high personalization.
\end{enumerate}

Input format: gender, question, response. Output Format: score: integer (1–5). Do not output explanations or any additional text.


\textbf{Sound event test set evaluation prompt}: The evaluation of the sound event test set is based on {matching}, using the following mapping relationships:


\begin{itemize}
  \item sigh : [sigh, ai, sighing, let out a sigh]
  \item cry : [cry, sad, heartbroken, weeping, tear up, sob, cried, melancholy]
  \item screaming : [shout, scream, yell, call out, cry out, loudly, shouting, terrified]
  \item laugh : [laugh, happy, funny, laughing out loud, laughter, smiling, giggling, amused, haha]
  \item throat clearing : [clear throat, clear voice, cleared throat]
  \item cough : [cough, coughing, coughed once, ill]
  \item sneeze : [sneeze, sneezing, sneezed, cold]
\end{itemize}


\textbf{Multi-label test set evaluation prompt}: You are a professional empathetic dialogue evaluation assistant. Your task is to {comprehensively evaluate} the quality of empathy in a system's response, based on the user's {paralinguistic information} (age, gender, emotional state, and sound events in the audio), and assign a score.

Empathetic dialogue system: A dialogue system capable of perceiving and adapting to the user’s paralinguistic features. It can recognize the user’s age, gender, current emotion, and sound events in the audio, and generate empathetic responses based on this information. The system should respond to all available paralinguistic cues whenever possible.

User Information: age, gender, emotional state, sound events in audio. The detailed evaluation dimensions are as follows.

\begin{itemize}
  \item Appropriateness of response content
  \begin{itemize}
    \item Does it understand and align with the user's semantics and background (e.g., age, gender)?
    \item Is it relevant, natural, and coherent in the context of dialogue?
    \item Does it address the core intent of the user’s utterance?
  \end{itemize}

  \item Effectiveness of emotional expression
  \begin{itemize}
    \item Does it correctly perceive and respond to the user’s emotional state?
    \item Is the emotional tone appropriate, non-offensive, and not misleading?
    \item Does it take into account the potential emotional impact of detected sound events?
  \end{itemize}
\end{itemize}

Scoring criteria range from 1 to 5 as follows:
\begin{itemize}
  \item[1.] The response is irrelevant, emotionally inappropriate, or potentially harmful.
  \item[2.] The response neglects or misinterprets the user’s paralinguistic features, showing a lack of empathy.
  \item[3.] There are noticeable issues in either content or emotional expression.
  \item[4.] The response is generally appropriate with minor omissions or imperfections.
  \item[5.] The response is natural, strongly aligned with paralinguistic features, and emotionally appropriate.
\end{itemize}

Only output a single Arabic numeral (1–5). Do not include explanations or any additional text.

\begin{table}[h]
\resizebox{\columnwidth}{!}{%
\begin{tabular}{@{}cccc@{}}
\toprule
\textbf{Task} & \textbf{Total Hours} & \textbf{Language} & \textbf{Open-sourced Dataset} \\ \midrule
\textbf{ASR} & 24k & EN, CN & \begin{tabular}[c]{@{}c@{}}Wenetspeech~\citep{wenetspeech}, \\ AISHELL-l~\citep{aishell1}, \\ AISHELL-2~\citep{aishell2}, \\ LibriSpeech~\citep{librispeech}\end{tabular} \\ \midrule
\textbf{VED} & 2.5k & CN & \begin{tabular}[c]{@{}c@{}}Audioset~\citep{jort_audioset_2017},\\ ESC-50~\citep{piczak2015dataset},\\ Vocal Sound~\citep{gong2022vocalsound},\\Nonspeech7k~\citep{rashid2023nonspeech7k},\\AISHELL-2~\citep{aishell2}\end{tabular} \\ \midrule
\textbf{SER} & 1k & EN, CN & \begin{tabular}[c]{@{}c@{}}ESD~\citep{zhou2021esd},\\ IEMOCAP~\citep{busso2008iemocap},\\ MSP-Podcast~\citep{martinez2020msp},\\ MER2023~\citep{Lian2023MER},\\ MELD~\citep{2019meld},\\ BIIC-Podcast~\citep{2023BIIC_Podcast}\end{tabular} \\ \midrule
\textbf{SGC} & 7.5k & EN, CN & \begin{tabular}[c]{@{}c@{}}KeSpeech~\citep{tang2021kespeech},\\  Datatang-Kid~\citep{datatang_page}, \\ Magicdata-Read~\citep{magicdata_read},\\ AISHELL-l~\citep{aishell1}, \\ AISHELL-2~\citep{aishell2}, \\LibriSpeech~\citep{librispeech},\\Kaggle-CommonVoice~\citep{kagglecv}\end{tabular} \\ \midrule
\textbf{SAP} & 6.8k & EN, CN & \begin{tabular}[c]{@{}c@{}}KeSpeech~\citep{tang2021kespeech},\\  Datatang-Kid~\citep{datatang_page}, \\ Magicdata-Read~\citep{magicdata_read}, \\ AISHELL-ASR0060~\citep{aishell_page}, \\ AISHELL-ASR0018~\citep{aishell_page},\\Kaggle-CommonVoice~\citep{kagglecv}\end{tabular} \\ 
 \bottomrule
\end{tabular}%
}
\caption{Details of the multi-task training data for OSUM-EChat.}
\label{tab:data_um}
\end{table}

\subsection{A.3 Training data}

\subsubsection{Speech understanding}
For the speech understanding task, we first adopted the OSUM~\cite{osum} dataset and leveraged the pre-trained OSUM understanding model to perform automatic label generation on existing single-label data—specifically, adding multi-dimensional labels such as emotion, style, speaker attributes, and events to each data entry, thereby constructing a dataset required for multi-task understanding. We incorporated the results of this ``automatic labeling" as weak supervision signals into the database, along with high-quality manually annotated samples. This strategy offers dual advantages: it enables the model to expand its task scope based on existing single-task data without the need for additional large-scale data collection, and it ensures the logical and standard consistency of the newly added labels by utilizing the prior knowledge already acquired by the model. Detailed specifics of the aforementioned data can be found in Table~\ref{tab:data_um}.

\subsubsection{Text Q\&A}
The construction of text question-and-answer data strictly adheres to the configuration standards of Alpaca-CoT~\cite{all_t2t} and encompasses multiple publicly available text Q\&A datasets. To better align the data with the characteristics of voice conversation tasks, we designed a targeted multi-level filtering mechanism: the core objective is to preserve the fundamental intelligence that the model requires as a voice conversation system. Consequently, specialized task data in the text domain—such as code parsing and complex data processing—are excluded, given their low relevance to voice interaction scenarios; excessive inclusion of such data could divert the model's training focus from core voice conversation capabilities. Additionally, all conversation entries containing special symbols (e.g., links, URLs, and email addresses) have been filtered out, as these may introduce redundant information irrelevant to the conversation, thereby impeding the model's ability to capture core semantics.

\subsubsection{Speech synthesis}
Our speech synthesis data primarily originates from Emilia~\cite{emilia} and LibriHeavy~\cite{libriheavy}, with specific processing methods as follows: The full dataset of Emilia is adopted to ensure the completeness and richness of the corpus, while 10,000 hours of content are selected from LibriHeavy to balance data scale and quality. Additionally, we specifically incorporated 2,000 hours of high-quality speech synthesis data from private studio recordings—this type of data is collected in a professional recording environment and strictly controlled in terms of pronunciation accuracy and intonation naturalness. The core purpose of introducing private data is to mitigate the problem of synthesis instability caused by minor inaccuracies in open-source data (such as pronunciation deviations and background noise), thereby improving the consistency and reliability of speech synthesis results.

\subsubsection{Audio Q\&A}
The first part generates voice conversation data by expanding ASR data, with the specific process as follows: Select open-source WenetSpeech~\cite{wenetspeech} Chinese ASR data (containing 10,000 hours) and Gigaspeech~\cite{gigaspeech}   English ASR data (containing 10,000 hours). First, input the text transcripts of these ASR data into the large model Qwen2.5-7B-instruct~\cite{llm:qwen2.5} to generate responses, and explicitly require the model to answer as a voice assistant through prompts—for example, the content should be concise and clear, avoiding lengthy and complex expressions to fit colloquial interaction scenarios. Next, adjust the prompts to control the Qwen2.5-7B-instruct model, treating the ASR transcripts as answers to specific questions, and then generate corresponding questions, thereby constructing complete conversation logic through this ``answer-to-question" approach. Subsequently, input the generated text conversation pairs into Qwen2.5-7B-instruct again for logical judgment and scoring, focusing on selecting samples with close semantic relevance and clear logical chains between questions and answers, while filtering out low-quality data with irrelevant answers or logical breaks. Finally, use Cosyvoice2~\cite{tts:cosyvoice2} combined with random audio prompts to synthesize audio corresponding to the text generated by Qwen2.5-7B-instruct, converting text conversations into audible voice interaction data through speech synthesis technology, and ultimately constructing approximately 15k hours of conversation data.

The second part of the data is voice conversation data converted from text Q\&A data. We conducted targeted expansion and optimization on the Text Q\&A data introduced earlier: Considering users' demand for information acquisition efficiency in voice interaction scenarios, we limited both questions and answers to within 150 words through a length filtering mechanism to ensure concise and focused content, avoiding reduced voice interaction experience due to overly long text. Meanwhile, we uniformly performed speech synthesis on the filtered question and answer texts, converting text-based Q\&A content into natural and fluent voice data, which not only retains the semantic integrity of the original text data but also endows it with auditory attributes suitable for voice interaction scenarios, ultimately forming 4k hours of Q\&A data.

The third part is our private real Q\&A voice data, with a scale of approximately 1.5k hours. Unlike the synthetic data generated by technical means in the first two parts, this part of the data comes from human conversation records in real scenarios, where both questions and answers are original audio materials without subsequent manual modification or machine-generated processing. Such data can accurately capture detailed features in real voice interactions, such as intonation changes, emotional colors, and colloquial expression habits, providing corpus support close to actual application scenarios for model training. It helps further improve the model's adaptability to real conversation scenarios and reduces the problem of ``disconnection between laboratory effects and practical applications" caused by training with purely synthetic data.

\section{B Training}
\subsection{B.1 Training details}
\label{appendix:training_details}
We completed the entire training of OSUM-EChat following the approach of ``understanding,  generation, and empathy." The following sections will provide a detailed description of each stage.

\subsubsection{Multi-task understanding training}
In this phase, we conduct joint training across multiple paralinguistic perception tasks, culminating in the development of a large-scale audio understanding model. The integrated tasks encompass five core functionalities: punctuated automatic speech recognition (ASR), emotion recognition, age estimation, gender identification, and sound event detection. Once the model has acquired paralinguistic recognition capabilities, we leverage the OSUM model to expand the labels of single-paralinguistic data, generating pseudo-labeled datasets that include all paralinguistic tags. This enables the model to simultaneously recognize all paralinguistic attributes, thereby establishing a foundational framework for subsequent integration with dialogue-related tasks.

Building on the experience of OSUM~\cite{osum}, we adopt an ASR+P training strategy, divided into four stages, with each stage trained for 1 epoch. In the first stage, only the speech adapter is unfrozen, and the ASR data is used for modality alignment training. In the second stage, both the speech adapter and speech encoder are unfrozen, and multi-task training is conducted following the ASR+P paradigm. In the third stage, only the speech encoder and speech adapter are unfrozen, and the ``only P" task is introduced. This task directly predicts the paralinguistic labels instead of the ASR results. Experiments show that the accuracy of the ``only P" task is similar to that of the ASR+P model and outperforms skipping the ASR+P stage and directly performing the ``only P" task. In the fourth stage, the ``only P" task is used to fill in missing labels, build full-label data, and integrate it into the training process, ultimately completing the full-label recognition task.

It is important to note that the data across the four stages is incrementally added to the model, and the training of each stage does not exclude data from previous stages, thereby maximizing the model’s comprehensive audio understanding capability.

\subsubsection{End-to-End s2s transition training}
This stage aims to equip the OSUM-based understanding model with the ability for speech-to-speech dialogue. To achieve this, we designed a two-step transition training process, with each training phase consisting of 2 epochs:

\textbf{T2S optimization}: First, we incorporate audio understanding task data with a weight of 0.1 to maintain the model's audio understanding capability. Next, we fine-tune the LLM using TTS task data, where the input to the LLM is text data and the output is speech tokens. We use Cross-Entropy Loss (CE Loss) to align the generated speech token sequence with the ground truth speech token sequence. After completing the first epoch, to prepare for the subsequent streaming speech dialogue tasks, we introduce the ``Interleaved TTS" task. In this task, the LLM’s input is still text, but the model alternates between text tokens and speech tokens in a 6:18 ratio, where text tokens represent the input text. During the second epoch, we randomly convert half of the normal TTS data to the ``Interleaved TTS" task. Since this phase modifies the parameters of the LLM, we also include text dialogue data to maintain its core text-based dialogue capabilities.

\textbf{S2S generation training}: Building on the first phase, we further conduct joint training on the end-to-end speech dialogue corpus. This phase uses a ``speech input → text \& speech output" supervision format. During the dialogue task training, the LLM receives the feature sequence of the question speech and outputs both text tokens and speech tokens. The speech-to-speech task is divided into two categories: non-streaming and streaming, with the main difference being the organization of the output tokens. The non-streaming task first outputs the text tokens, followed by the speech tokens. The streaming task alternates between text and speech tokens, maintaining the 6:18 ratio to ensure stable speech synthesis. Following the approach of CosyVoice2~\cite{tts:cosyvoice2}, we introduce an End-of-Text (EOS) symbol, which indicates that the text generation is complete, after which only speech tokens are generated, avoiding excessive padding values during the text generation phase and thus saving resources. This phase’s training, similar to the previous one, is divided into two epochs: the first epoch trains the non-streaming task, and the second epoch introduces the streaming task, during which half of the speech dialogue data is randomly switched to the streaming format. Meanwhile, we continue to use the complete TTS data, text data, and 0.1 times the understanding task data to maintain the capabilities of other tasks.

\subsubsection{Empathy training}
 To enable the effective transfer of paralinguistic information in dialogue generation, we employ the ``\textit{linguistic-paralinguistic Dual Think}" strategy. During answer generation, the model first perceives both the textual and paralinguistic information in the input audio, then integrates this information to provide a more coherent response. Building upon this, we further trained the previous model using the \textit{EChat-200K} dataset and incorporated non-paralinguistic question-answer data from the S2S stage. A weight of 0.2 was applied to emphasize the paralinguistic dialogue tasks, while the data for other tasks remained consistent with the previous phase.

Given that the \textit{EChat-200K} dataset contains only a small amount of fully-labeled speech Q\&A data, with most data containing only one paralinguistic label and non-paralinguistic dialogue data lacking task labels, it became a challenge to unify these data for learning the ``Think" capability. To address this challenge, we propose a ``Structured Sparse Learning" strategy, where all data is organized in a ``Think" format that includes all paralinguistic labels. For data with unknown labels, we use the ``unk" label as a placeholder. During training, the positions of the ``unk" labels are masked, preventing them from contributing to the backpropagation of the loss. Only labeled data will guide the learning of paralinguistic label placements. With this strategy, the model learns first to output linguistic information and then sequentially output different parainguistic expressions, thus effectively integrating various types of training data. This phase is trained for one epoch.

Once the model has acquired the ability to generate comprehensive responses to paralinguistic-rich inputs, we draw inspiration from the Chain-of-Thought (COT) approach. We continue using the \textit{EChat-200K} dataset and introduce direct speech dialogue tasks, training for one epoch. Through self-distillation, the model achieves paralinguistic question-answering capabilities comparable to those of the model with the ``Think" stage, even without the inclusion of the Think phase.

\begin{table}[h!]
\centering
\footnotesize
\centering
\resizebox{\columnwidth}{!}{%
\begin{tabular}{@{}c c c c c@{}} 
\toprule
\textbf{Task} & \textbf{Model} & \textbf{Metric} & \textbf{Test Set} & \textbf{Result} \\ \midrule
\multirow{12}{*}{\textbf{ASR}} 
& Whisper-L-v3 & \multirow{12}{*}{\begin{tabular}[c]{@{}c@{}}WER/CER\\ (${\%}$, ${\downarrow}$)\end{tabular}} & \multirow{12}{*}{\begin{tabular}[c]{@{}c@{}}\textbf{WenetSpeech}\\ test-net/test-meeting\\ \textbf{AISHELL2}\\ mic/ios/android \\ \textbf{Librispeech}\\ test-clean/test-other\end{tabular}} & \begin{tabular}[c]{@{}c@{}}10.48/18.87\\ -/4.96/- \\ 1.82/3.50\end{tabular} \\ \cmidrule(lr){2-2} \cmidrule(lr){5-5} 


& Qwen2-Audio &  &  & \begin{tabular}[c]{@{}c@{}}-/-\\ 3.0/3.0/2.9 \\ \textbf{1.6}/\textbf{3.6}\end{tabular} \\ \cmidrule(lr){2-2} \cmidrule(lr){5-5} 

& OSUM &  &  & \begin{tabular}[c]{@{}c@{}}6.46/\textbf{5.34} \\ \textbf{2.81}/\textbf{2.75}/\textbf{2.73} \\2.19/5.53\end{tabular} \\ \cmidrule(lr){2-2} \cmidrule(lr){5-5} 
& OSUM-EChat &  &  & \begin{tabular}[c]{@{}c@{}}6.47/9.90 \\2.89/3.04/2.89\\2.46/5.93\end{tabular} \\ \midrule

\multirow{6}{*}{\textbf{Event}} 
& Qwen2-Audio & \multirow{6}{*}{\begin{tabular}[c]{@{}c@{}}ACC\\ (${\%}$, ${\uparrow}$)\end{tabular}} & \multirow{6}{*}{\begin{tabular}[c]{@{}c@{}}\\ \textbf{VocalSound}\\ \textbf{MMAU (Sound)}\end{tabular}} & \begin{tabular}[c]{@{}c@{}} 93.3\\ 54.95\end{tabular} \\ \cmidrule(lr){2-2} \cmidrule(lr){5-5} 
& OSUM &  &  & \begin{tabular}[c]{@{}c@{}}82.58\\ -\end{tabular} \\ \cmidrule(lr){2-2} \cmidrule(lr){5-5} 
& OSUM-EChat &  &  & \begin{tabular}[c]{@{}c@{}} 82.58\\ 45.05\end{tabular} \\ \midrule

\multirow{6}{*}{\textbf{Emotion}} 
& Qwen2-Audio & \multirow{6}{*}{\begin{tabular}[c]{@{}c@{}}ACC\\ (${\%}$, ${\uparrow}$)\end{tabular}} & \multirow{6}{*}{\begin{tabular}[c]{@{}c@{}}\textbf{MELD}\\ test\\ \textbf{MER2023}\\ test\end{tabular}} & \begin{tabular}[c]{@{}c@{}}55.3\\ -\end{tabular} \\ \cmidrule(lr){2-2} \cmidrule(lr){5-5} 
& OSUM &  &  & \begin{tabular}[c]{@{}c@{}}53.38\\ \textbf{86.43}\end{tabular} \\ \cmidrule(lr){2-2} \cmidrule(lr){5-5} 
& OSUM-EChat &  &  & \begin{tabular}[c]{@{}c@{}}51.23\\ 85.93\end{tabular} \\ \midrule

\multirow{6}{*}{\textbf{Gender}} 
& Qwen2-Audio & \multirow{6}{*}{\begin{tabular}[c]{@{}c@{}}ACC\\ (${\%}$, ${\uparrow}$)\end{tabular}} & \multirow{6}{*}{\begin{tabular}[c]{@{}c@{}}\textbf{AISHELL-1}\\ test\\ \textbf{Kaggle-CommonVoice}\\ valid-test\end{tabular}} & \begin{tabular}[c]{@{}c@{}}97.36\\ 97.25\end{tabular} \\ \cmidrule(lr){2-2} \cmidrule(lr){5-5} 
& OSUM &  &  & \begin{tabular}[c]{@{}c@{}}\textbf{100}\\ 99.41\end{tabular} \\ \cmidrule(lr){2-2} \cmidrule(lr){5-5} 
& OSUM-EChat &  &  & \begin{tabular}[c]{@{}c@{}}\textbf{100}\\ \textbf{99.67}\end{tabular} \\ \midrule

\multirow{4}{*}{\textbf{Age}} 
& Qwen2-Audio & \multirow{4}{*}{\begin{tabular}[c]{@{}c@{}}ACC\\ (${\%}$, ${\uparrow}$)\end{tabular}} & \multirow{4}{*}{\begin{tabular}[c]{@{}c@{}}\textbf{Kaggle-CommonVoice}\\ valid-test\end{tabular}} & 35.53 \\ \cmidrule(lr){2-2} \cmidrule(lr){5-5} 
& OSUM &  &  & 76.52\\ \cmidrule(lr){2-2} \cmidrule(lr){5-5} 
& OSUM-EChat &  &  & \textbf{83.77} \\ \bottomrule
\end{tabular}%
}
\caption{Evaluation results of multi-tasking on public test sets.}
\label{tab:res_um}
\end{table}


\begin{table*}[t!]
\centering
\resizebox{0.95\linewidth}{!}{
\begin{tabular}{lcccccccc}
\toprule
\textbf{model} & \textbf{FLUB} & \textbf{gsm8k} & \textbf{logitQA} & \textbf{math23k} & \textbf{math401} & \textbf{ceval} & \textbf{mmlu} & \textbf{ifeval} \\
\midrule
Qwen2.5-3B-instruct & 51.92 & 85.14 & 44.29 & 85.90 & 67.58 & 73.58 & 62.28 & 65.18 \\
OSUM-EChat & 49.28 & 33.74 & 41.02 & 30.50 & 46.00 & 55.81 & 60.75 & 31.83 \\
\bottomrule
\end{tabular}
}
\caption{Comparison of text dialogue performance between OSUM-EChat and its LLM backbone across different benchmarks.}
\label{tab:txt_res}
\end{table*}

\begin{table*}[t]
\centering
\scriptsize
\resizebox{0.9\textwidth}{!}{ 
\begin{tabular}{@{}lcccccccc@{}}
\toprule
\textbf{Model} & \textbf{\makecell{Alpaca\\Eval}} & \textbf{\makecell{Common\\Eval}} & \textbf{\makecell{SD-\\QA}} & \textbf{\makecell{MMSU}} & \textbf{\makecell{OpenBook\\QA}} & \textbf{\makecell{IFEval}} & \textbf{\makecell{Adv\\Bench}} & \textbf{\makecell{Overall}} \\ \midrule
GPT-4o-Audio & 4.78 & 4.49 & 75.5 & 80.25 & 89.23 & 76.02 & 98.65 & 86.42 \\ 
Baichuan-Audio & 4.41 & 4.08 & 45.84 & 53.19 & 71.65 & 50.31 & 99.42 & 70.03 \\ 
GLM-4-Voice & 3.97 & 3.42 & 36.98 & 39.75 & 53.41 & 25.92 & 88.08 & 55.99 \\ 
SLAM-Omni & 1.9 & 1.79 & 4.16 & 26.06 & 25.27 & 13.38 & 94.23 & 33.84 \\ 
Mini-Omni2 & 2.32 & 2.18 & 9.31 & 24.27 & 26.59 & 11.56 & 57.5 & 31.32 \\ 
Mini-Omni & 1.95 & 2.02 & 13.92 & 24.69 & 26.59 & 13.58 & 37.12 & 27.9 \\ 
Moshi & 2.01 & 1.6 & 15.64 & 24.04 & 25.93 & 10.12 & 44.23 & 27.47 \\ 
 OSUM-EChat  & 4.29 & 3.39 & 24.41 & 60.11 & 78.46 & 32.36 & 92.88 & 63.12 \\ 
\bottomrule
\end{tabular}
}
\caption{Performance on Voicebench Benchmarks}
\label{tab:voicebench}
\end{table*}

\subsection{B.2 Design of prompt}

OSUM-EChat is an empathetic dialogue model with multi-task capabilities, and we use different system prompts to distinguish between tasks. Below is a detailed introduction to the system prompt framework we use. We adopt the following prompt format:

\[
\begin{array}{l}
\texttt{<|im\_start|>system} \\
\texttt{system prompt <|im\_end|>} \\
\texttt{<|im\_start|>user} \\
\texttt{[user prompt] <speech> <|im\_end|>} \\
\texttt{<|im\_start|>assistant} \\
\texttt{reply content} \\
\end{array}
\]

User prompts are only used in the speech understanding task to instruct the model on the understanding task to perform. For the system prompt, we use the following content:

\textbf{Speech dialogue task}: You are OSUM-chat, a speech-to-speech dialogue model. You understand both the meaning and paralinguistic cues in speech, and then respond with appropriate text and emotionally matching synthetic speech.

\textbf{Spoken dialogue task with think phase}:
You are OSUM-chat, a speech-to-speech dialogue model. You understand both the meaning and paralinguistic cues in speech. Before responding, first output your reasoning inside \texttt{<think>...</think end>}, analyzing the user’s words and vocal cues. Then generate a reply with appropriate text and emotionally matched synthetic speech.

\textbf{Spoken dialogue task with text-speech interleaved output}: You are OSUM-chat, a speech-to-speech dialogue model. You analyze speech (content + paralinguistic cues) and respond with interleaved text and emotionally-matched synthetic speech.

\textbf{Spoken dialogue task with text-speech interleaved output and think phase}: You are OSUM-chat, a speech-to-speech dialogue model. You analyze speech (both content and paralinguistic cues). Before responding, output your reasoning in \texttt{<think>...</think end>}. Then reply with interleaved text and emotionally matched synthetic speech.

\textbf{Speech-to-Text dialogue task}: You are OSUM-chat, a speech-to-text dialogue model. You understand both the meaning and paralinguistic cues in speech, and then respond exclusively with appropriate text.

\textbf{Text-to-Text dialogue task}: You are OSUM-chat, a text-to-text dialogue model. You understand user input in text and then respond exclusively with appropriate text.

\textbf{Speech understanding task}: You are OSUM-chat, an audio understanding model. You can transcribe speech accurately and analyze paralinguistic cues to provide precise text responses.

\textbf{Speech synthesis task}: You are OSUM-chat, a speech synthesis model. You generate natural and fluent speech from text input.

\section{C Results}
\subsection{C.1 Speech understanding}
We evaluated the performance of OSUM-EChat on more open-source test sets, and the specific results are shown in Table~\ref{tab:res_um}.

\subsection{C.2 Text Q\&A}
To verify the intelligence retention of OSUM-EChat in text-to-text tasks, we conducted comprehensive reasoning on multiple public text test sets and compared it with the Qwen2.5-3B-instruct model. Details are shown in Table~\ref{tab:txt_res}.

\subsection{C.3 Audio Q\&A}
We also conducted speech-to-text inference on Voicebench~\cite{voicebench} and compared it with mainstream speech response models. As shown in Table~\ref{tab:voicebench}, OSUM-EChat demonstrates comparable performance to mainstream models in the speech-to-text dialogue task.

\subsection{C.4 Paralinguistic instruction-following}
OSUM-EChat is designed to implicitly model paralinguistic information and integrate such information into its responses in an ``implicit" manner, rather than describing it directly. To assess OSUM-EChat's intuitive perception of paralinguistics, we conducted a paralinguistic instruction-following test.

In this test, we constructed 200 test samples for each of the four paralinguistic categories (i.e., emotion, age, gender, and caption), with specific designs as follows:
\begin{itemize}
    \item Emotion test set: Instructions require the model to repeat a given sentence with a specified emotion;
    \item Age and gender test sets: Users directly inquire about their own age or gender, and the model is required to provide corresponding answers by analyzing the paralinguistic information in the user's speech;
    \item Sound event test set: The model is tasked with identifying the category of sound events input synchronously with the speech.
\end{itemize}

The evaluation was still conducted using ChatGPT-4o~\cite{gpt4o}, and emotion labels were extracted using emotion2vec-large~\cite{emotion2vec}.

The specific test results are presented in Table~\ref{tab:res_paralinguistic_performance}.


\begin{table}[h!]
    \centering
    \begin{tabular}{lcccc}
        \toprule
                        models  & Emotion & Age  & Gender & Event   \\
        \midrule
        qwen2.5-omni       & 39.5   & -    & -      & 37      \\
        freeze-omni        & 41.2   & -    & -      & -       \\
        GLM-4-voice        & 58.5   & -    & -      & -       \\
        baichuan-audio     & 38.8   & -    & -      & -       \\
        kimi-audio         & 58.5   & 34   & 62     & 12      \\
        opens2s            & 41.5    & 30   & 37.5   & -       \\
        OSUM-EChat         & 83.2   & 70.0 & 95.0   & 76.3  \\
        \bottomrule
    \end{tabular}
    \caption{Performance of Different Models on Paralinguistic-related Instruction-following Test Sets. - Indicates that the model lacks this capability.
}
    \label{tab:res_paralinguistic_performance}
\end{table}

\section{D Ethical Statement } 

To achieve empathetic dialogue, the OSUM-EChat system, developed for research purposes, captures and analyzes paralinguistic information from users' speech, including emotion, gender, and age. Such data processing carries inherent risks of privacy infringement, which necessitates rigorous ethical safeguards.

This study strictly adheres to established ethical guidelines for human subjects research. All audio-derived user information is exclusively utilized for the development and evaluation of empathetic dialogue capabilities, with no secondary use or disclosure for unrelated research objectives or commercial applications. Before data collection, participants are provided with a comprehensive explanation of the purpose, scope, and processing methods of audio analysis, and written informed consent is obtained. Data processing is conducted only with explicit participant authorization, ensuring full transparency and preserving participants’ rights to information, access, and withdrawal. These measures are implemented to minimize potential risks of privacy violations and ensure compliance with ethical standards.










\end{document}